\documentclass[twocolumn]{aastex63}

\usepackage{amsmath}

\usepackage{float}

\makeatletter
\newcommand\DefineObj[3][\@empty]{%
  \expandafter\newcommand\csname pkgwobj@l#2\endcsname{#3}%
  \ifx\@empty#1%
    \expandafter\newcommand\csname pkgwobj@s#2\endcsname{#3}%
  \else%
    \expandafter\newcommand\csname pkgwobj@s#2\endcsname{#1}%
  \fi}%
\newcommand{\pkgw@printobj@long}[1]{%
  \expandafter\ifx\csname pkgwobj@l#1\endcsname\relax%
    \textbf{[unknown object!]}%
  \else%
    \csname pkgwobj@l#1\endcsname%
  \fi}%
\newcommand{\pkgw@printobj@short}[1]{%
  \expandafter\ifx\csname pkgwobj@l#1\endcsname\relax%
    \textbf{[unknown object!]}%
  \else%
    \csname pkgwobj@s#1\endcsname%
  \fi}%
\newcommand{\obj}{\@ifstar{\pkgw@printobj@long}{\pkgw@printobj@short}}%
\makeatother
\DefineObj[2MASS~J2139$+$02]{2m2139}{2MASS J21392676$+$0220226}

\DefineObj[SIMP~J0136$+$09]{s0136}{SIMP J01365662+0933473}

\DefineObj[2MASS~2224--01]{2m2224}{2MASSW J2224438-015852}

\shorttitle{Patchy Forsterite Clouds}
\shortauthors{Vos et al.}
\graphicspath{{./}{figures/}}

\begin{document}

\title{Patchy Forsterite Clouds in the Atmospheres of Two Highly Variable Exoplanet Analogs}

\correspondingauthor{Johanna M. Vos}
\email{jvos@amnh.org}

\author[0000-0003-0489-1528]{Johanna M. Vos}
\affiliation{Department of Astrophysics, American Museum of Natural History, New York, NY 10024, USA}

\author[0000-0003-4600-5627]{Ben Burningham}
\affiliation{Centre for Astrophysics Research, Department of Physics, Astronomy and Mathematics, University of Hertfordshire, Hatfield AL10 9AB, UK}

\author[0000-0001-6251-0573]{Jacqueline K. Faherty}
\affiliation{Department of Astrophysics, American Museum of Natural History, New York, NY 10024, USA}

\author[0000-0003-0548-0093]{Sherelyn Alejandro}
\affiliation{Hunter College, City University of New York, 695 Park Avenue, New York, NY 10065, USA}
\affiliation{Department of Astrophysics, American Museum of Natural History, New York, NY 10024, USA}

\author[0000-0003-4636-6676]{Eileen Gonzales}
\affiliation{Department of Astronomy and Carl Sagan Institute, Cornell University, 122 Sciences Drive, Ithaca, NY 14853, USA}
\altaffiliation{51 Pegasi b Postdoctoral Fellow}

\author[0000-0002-2682-0790]{Emily Calamari}
\affiliation{The Graduate Center, City University of New York,
New York, NY 10016, USA}
\affiliation{Department of Astrophysics, American Museum of Natural History, New York, NY 10024, USA}

\author[0000-0001-8170-7072]{Daniella Bardalez Gagliuffi}
\affiliation{Department of Physics \& Astronomy, Amherst College, 25 East Drive, Amherst, MA 01003, USA}
\affiliation{Department of Astrophysics, American Museum of Natural History, New York, NY 10024, USA}

\author[0000-0001-6627-6067]{Channon Visscher}
\affiliation{Chemistry \& Planetary Sciences, Dordt University, Sioux Center, IA, USA}
\affiliation{Center for Extrasolar Planetary Systems, Space Science Institute, Boulder, CO, USA}

\author[0000-0003-2278-6932]{Xianyu Tan}
\affiliation{Atmospheric, Ocean, and Planetary Physics, Department of Physics, University of Oxford, UK}

\author[0000-0002-4404-0456]{Caroline V. Morley}
\affiliation{Department of Astronomy, University of Texas at Austin, Austin, TX 78712, USA}

\author[0000-0002-5251-2943]{Mark Marley}
\affiliation{Department of Planetary Sciences and Lunar and Planetary Laboratory, University of Arizona, Tucson, AZ, USA}

\author[0000-0002-8871-773X]{Marina E. Gemma}
\affiliation{Department of Earth and Environmental Sciences, Columbia University, New York, NY 10027, USA}
\affiliation{Department of Earth and Planetary Sciences, American Museum of Natural History, New York, NY 10024, USA}

\author[0000-0001-8818-1544]{Niall Whiteford}
\affiliation{Department of Astrophysics, American Museum of Natural History, New York, NY 10024, USA}

\author{Josefine Gaarn}
\affiliation{Centre for Astrophysics Research, Department of Physics, Astronomy and Mathematics, University of Hertfordshire, Hatfield AL10 9AB, UK}

\author{Grace Park}
\affiliation{Palisades Park High School, Palisades Park, NJ 07650, USA}



\begin{abstract}
We present an atmospheric retrieval analysis of a pair of highly variable, $\sim200~$Myr old, early-T type planetary-mass exoplanet analogs \obj*{s0136} and \obj*{2m2139} using the \textit{\textit{Brewster}} retrieval framework. Our analysis, which makes use of archival $1-15~\mu$m spectra, finds almost identical atmospheres for both objects. 
For both targets, we find that the data is best described by a patchy, high-altitude forsterite (Mg$_2$SiO$_4$) cloud above a deeper, optically thick iron (Fe) cloud. Our model constrains the cloud properties well, including the cloud locations and cloud particle sizes.  We find that the patchy forsterite slab cloud inferred from our retrieval may be responsible for the spectral behavior of the observed variability. Our retrieved cloud structure is consistent with the atmospheric structure previously inferred from spectroscopic variability measurements, but clarifies this picture significantly. We find consistent C/O ratios for both objects which supports their formation within the same molecular cloud in the Carina-Near Moving Group. Finally, we note some differences in the constrained abundances of H$_2$O and CO which may be caused by data quality and/or astrophysical processes such as auroral activity and their differing rotation rates. The results presented in this work provide a promising preview of the detail with which we will characterize extrasolar atmospheres with JWST, which will yield higher quality spectra across a wider wavelength range.  


\end{abstract}



\section{Introduction} \label{sec:intro}
 

Since 1995, over 5,000 planets orbiting other stars have been discovered using a variety of techniques. Of particular interest are the small number of imaged exoplanets that offer the most direct view into their atmospheres \citep[e.g.][]{,Chauvin2004,Marois2008, Lagrange2010, Macintosh2015, Bohn2020, Bohn2021}. Current and forthcoming telescopes such as the James Webb Space Telescope (JWST) and 30-metre telescopes (e.g. the Extremely Large Telescope, ELT; first light expected $\sim2027$) will enable direct exoplanet studies \citep[e.g.][]{Carter2022}, but interpretation of these results will hinge on a thorough understanding of their complex atmospheres. In the same time frame, a large number of brown dwarfs have been discovered that overlap in mass, age and temperature with the directly-imaged exoplanets \citep{Faherty2016, Liu2016}. These isolated exoplanet analogs are much easier to study without the glare of a bright host star, and offer crucial insight into the nature of extrasolar atmospheres.

Condensate clouds have emerged as one of the most confounding issues for attempts to characterize extrasolar atmospheres in detail. These clouds are predicted to form in atmospheres with temperatures below $\sim2200~$K \citep{Burrows1997,Ackerman2001, Helling2008, Allard2012}, and dramatically shape the observed appearance of brown dwarfs and directly imaged exoplanets. For example, the red near-IR colors of L dwarfs are thought to be caused by the presence of silicate clouds while the blue near-IR colors of T dwarfs are caused by those same clouds sinking below the photosphere \citep{Ackerman2001, Saumon2008}. 

 Variability monitoring is a powerful probe of  extrasolar atmospheres.
 Solar  system  gas  giants,  giant  exoplanets,  and  brown  dwarfs  exhibit temporal variability in broadband emission and spectra as they rotate. This observed photometric and spectroscopic variability is generally thought to be driven by the presence of inhomogeneous clouds.  In the case of Jupiter, unresolved observations at $5~\mu$m reveal periodic variability with amplitudes exceeding 20\% \citep{Gelino2000, Ge2019}, caused by a range of inhomogeneous atmospheric features such as the Great Red Spot, the North Equatorial Belt and the North Temperate Belt \citep{Ge2019}. Beyond the solar system, these kind of observations have been carried out for a large number of field brown dwarfs \citep{Buenzli2014, Radigan2014, Metchev2015}, isolated exoplanet analogs \citep{Biller2015, Lew2016, Biller2018, Vos2019, Vos2022} and a smaller sample of wide orbit companions \citep{Zhou2016,Zhou2019,Zhou2020}. Large surveys have shown that variability is common \citep{Radigan2014, Metchev2015}, particularly among the young, low-gravity sample \citep{Vos2019, Vos2022}. In-depth spectroscopic monitoring studies, many of which make use of the WFC3 instrument on the Hubble Space Telescope, have also revealed clues on the vertical cloud structure in extrasolar atmospheres \citep[e.g.][]{Apai2013, Buenzli2015, Lew2016}, with wavelength dependent amplitudes and phase shifts suggesting the presence of regions of thin and thick cloud cover.
 
Atmospheric retrieval algorithms have also emerged as a powerful probe of extrasolar atmospheres, complementary to the information that can be gleaned from variability monitoring observations. 
Originally developed for remote sensing solar system atmospheres, atmospheric retrievals have recently been used to study exoplanet and brown dwarf atmospheres \citep[e.g.][]{Madhusudhan2009,Lee2012, Line2014, MacDonald2017, Burningham2017, Burningham2021, Molliere2019, Molliere2020, Kitzmann2020,Barstow2020}. 
In comparison to traditional approaches, whereby spectra of brown dwarfs and exoplanets are interpreted via comparison with pre-computed grids of self-consistent model atmospheres, atmospheric retrievals allow for a more efficient exploration of the parameter space with fewer initial assumptions. This data-driven technique allows for empirical constraints to be placed directly on key atmospheric parameters. 

The \textit{\textit{Brewster}} retrieval code \citep{Burningham2017} was specifically designed to capture the complexity of clouds in giant extrasolar atmospheres, and its effectiveness has been demonstrated across the L and T spectral sequence \citep{Burningham2017, Burningham2021, Gonzales2020, Gonzales2021, Calamari2022, Gaarn2022}. \citet{Burningham2021} in particular demonstrated the power of this algorithm when combined with wide wavelength spectroscopy, finding that the atmosphere of the mid-L dwarf \obj*{2m2224} (\obj{2m2224} hereafter) is best described by a rather complex model that includes enstatite (MgSiO$_3$) and quartz (SiO$_2$) cloud layers at shallow pressures, combined with a deeper iron (Fe) cloud deck. 

To date, the techniques of time-resolved variability monitoring and atmospheric retrievals have been applied separately to reveal extrasolar atmospheres. In this work, we present the first attempt to combine the information gleaned from each technique by performing atmospheric retrievals on a pair of exoplanet analogs whose variability properties have been studied in detail. By combining results gained from our retrievals with past studies of the time-resolved behavior of their atmospheres, our goal is to provide the clearest view to date of two extrasolar atmospheres.

\section{Twin Variable Worlds: SIMP  J0136+09 and 2MASS J2139+02} \label{sec:targets}


The T2.5 dwarf \obj*{s0136} (\obj{s0136} hereafter) was the first brown dwarf known to display repeatable, periodic variability due to inhomogeneous condensate clouds. The initial discovery paper by \citet{Artigau2009} revealed a rotation period of $\sim2.4~$hr, a peak-to-peak $J$-band amplitude of $\sim50$ mmag and significant light curve evolution between consecutive nights. 
\citet{Yang2016} refined the rotation period measurement to $2.414 \pm 0.078~$hr based on photometric monitoring with the Spitzer Space Telescope. 
\citet{Vos2017} determined that \obj{s0136} is inclined close to equator-on  ($i=80^{+10~\circ}_{-12}$) based on its rotation period and a $v\sin i$ measurement.
\citet{Kao2016} detected circularly polarized emission in the $4-8~$GHz band consistent with auroral activity, and estimated a magnetic field strength of $>2.5$~kG.
\citet{Gagne2017} identified \obj{s0136} as a member of the $\sim200~$Myr Carina-Near moving group, and estimate a mass of $12.7\pm1.0~M_{\mathrm{Jup}}$, placing it at the nominal boundary between planets and brown dwarfs.

The T1.5 dwarf \obj*{2m2139} (\obj{2m2139} hereafter) has emerged as a twin to \obj{s0136}, sharing many observable and physical properties. 
\citet{Radigan2012} first reported extremely high-amplitude near-IR variability in \obj{2m2139}. Using multiple epochs of ground-based monitoring observations, they measured peak-to-peak amplitudes as high as $26\%$ in $J$-band and a period of $7.721\pm0.005~$hr. This extremely high amplitude is matched only by VHS~J125601.92--125723.9~B \citep{Bowler2020}, a late-type L dwarf co-moving with a young late-M binary system, and thus represents a strong outlier in the population of known variables.
By examining the observed amplitudes across different bands, \citet{Radigan2012} attributed the observed variability to be due to patches of thin and thick clouds. \citet{Vos2017} determined that \obj{2m2139} is inclined equator-on  by combining its rotation period with a $v\sin i$ measurement. A new parallax measurement for \obj{2m2139} reported by \citet{Zhang2021} {revealed} that \obj{2m2139} is also a member of the Carina-Near moving group, making it a true sibling of \obj{s0136}.

As the first two high-amplitude L/T transition variables known, \obj{2m2139} and \obj{s0136} have often been studied side-by-side. \citet{Apai2013} followed up both objects with spectroscopic variability monitoring using the Hubble Space Telescope (HST) WFC3 grism mode. These observations revealed spectroscopic variations that were remarkably similar in both targets - $J$ and $H$ band brightness variations with minimal color and spectral changes. The authors find that the observed color changes and spectral variations in both cases are best described by an atmosphere with a heterogeneous mixture of cool, thick clouds and warmer, thin clouds, as opposed to patches of cloudy and clear patches. The authors estimate a $\sim300$ K temperature difference between the two cloud regions for \obj{s0136} and \obj{2m2139}. More recently, \citet{Apai2017} obtained long-term monitoring of both objects using the Spitzer Space Telescope, finding that they both exhibit similar long-term light curve evolution. They find that the long-term behavior of their light curves can be explained by sinusoidally modulated zonal bands as well as a bright spot.

With their recent identifications as members of the Carina-Near association, the comparison of \obj{2m2139} and \obj{s0136} is even more fitting today.  With similar ages, masses and temperatures, this unique pair of objects allows us to examine the atmospheres of two objects that presumably share a formation history and subsequent evolution. In this paper, we employ the \textit{Brewster} spectral retrieval framework to compare and contrast the constituents of their atmospheres. 

\section{Spectral Data}\label{sec:spectra}

In order to gain the most comprehensive view of the atmospheres of \obj{s0136} and \obj{2m2139}, we use the widest wavelength range available for both targets. 
For \obj{s0136}, our spectrum is comprised of a $1-2.5~\mu$m IRTF/SpeX Prism spectrum \citep[$R\sim120$, S/N$\sim180$;][]{Burgasser2008b}, a $2.5-5~\mu$m AKARI/IRC spectrum reported by \citet{Sorahana2013} ($R\sim100$, S/N$\sim9$) and a $5-15~\mu$m Spitzer/IRS spectrum presented by \citet{Filippazzo2015} ($R\sim90$, S/N$\sim15$). Our \obj{2m2139} spectrum consists of a $1-2.5~\mu$m IRTF/SpeX spectrum from \citet{Burgasser2006} ($R\sim120$, S/N$\sim170$) and a $5-15~\mu$m Spitzer/IRS spectrum presented by \citet{Suarez2022} ($R\sim90$, S/N$\sim6$). The near-infrared data for both sources was flux calibrated using  their parallax and 2MASS $J$ photometry, while the AKARI/IRC and Spitzer/IRS spectra were already flux calibrated \citep{Filippazzo2015, Suarez2022}. We note that for \obj{2m2139}, while \citet{Suarez2022} corrected their Spitzer/IRS spectrum flux calibration to match \textit{WISE} $W3$-band photometry, we use the IRS instrumental flux calibration for this work.

\begin{table*}
\begin{center}
\begin{tabular}{llll}
Property                & \obj{2m2139}                      &\obj{s0136}         & Reference \\\hline \hline
RA                      & 21:39:26.76936                    & 01:36:57        &  C03     \\
Dec                     &+02:20:22.6968                     & 09:33:47              &  C03   \\
$\mu\alpha$ cos $\delta$ (mas yr$^{-1}$)&  $485.9\pm2.0$    & $1238.982\pm1.189$    & C03, G18    \\
$\mu\delta$ (mas yr$^{-1}$) &  $124.8\pm2.7$                & $-17.353 \pm 0.841$     & S13, G18  \\
RV (km s$^{-1}$)               &  $-25.1\pm0.3$             & $12.3 \pm 0.8$         & V17           \\
Spectral Type  (Opt/IR) &       T2/T1.5      &  T2/T2.5  & P16/B06,     P16/A06       \\
Parallax (mas)    &  $96.5\pm1.1$                     & $	163.7\pm0.7$         &  Z21     \\
Rotation Period (hr)    & $7.619 \pm 0.0168 $               & $2.414\pm0.078$    & Y16         \\
$v\sin i$ (km s$^{-1}$)  & $18.7 \pm 0.3$                    & $52.8^{+1.0}_{-1.1} $   & V17       \\ \hline
Photometry \\ \hline
2MASS $J$ (mag) &    $15.264\pm0.049$                   & $13.455\pm0.030$   &   C03     \\\hline
Fundamental Parameters from SED Analysis \\ \hline
$L_{bol}$                   &    $-4.84\pm0.02$.        &   $-4.65 \pm0.06$   &   This work         \\
$T_{\mathrm{eff}}$ (K)      &   $1040\pm60$             &   $1150\pm70$  &   This work         \\
Radius $(R_{\mathrm{Jup}})$ &   $1.14\pm0.11$             &  $1.15\pm0.11$  &   This work         \\
Mass  $(M_{\mathrm{Jup}})$  &  $16.2\pm8.8$             &   $17.8\pm11.9$   &     This work       \\
$\log g$                    & $4.5\pm0.4$               &    $4.5\pm0.4$ &   This work        \\ \hline
Fundamental Parameters from Retrieval Analysis \\ \hline
$L_{bol}$                   &  $-4.91\pm0.005$          &   $-4.71\pm0.002$   &   This work     \\
$T_{\mathrm{eff}}$ (K)$^{a}$      & $1360\pm20$               &   $1329\pm17$  &    This work     \\
Radius $(R_{\mathrm{Jup}})^{a}$ &$0.61\pm0.02$              & $0.81\pm0.03$  &   This work      \\
Mass  $(M_{\mathrm{Jup}})^{a}$     & $2.82\pm1.0$              &   $4.71\pm0.8$   &    This work     \\
$\log g$                    &$4.27\pm0.2$               &    $4.25\pm0.08$ &    This work    \\ \hline

\end{tabular}
\end{center}
\caption{Observed properties and derived fundamental parameters for \obj{2m2139} and \obj{s0136} }
\tablerefs{C03: \citet{Cutri2003}; G18: \citet{Gaia2018};  S13: \citet{Smart2013}; V17: \citet{Vos2017}; Y16: \citet{Yang2016}; Z21: \citet{Zhang2021}; P16: \citet{Pineda2016}; B06: \citet{Burgasser2006}; A06: \citet{Artigau2006}.}
\tablecomments{ $^{a}$ We find that some of the fundamental parameters derived from our retrieval show poor consistency with the SED-derived values.  We discuss these discrepancies in Section \ref{sec:SED_comparison}.}
\label{tab:props}
\end{table*}

\section{Fundamental Parameters via SEDkit}
We analyze the spectral energy distributions (SEDs) of our two targets to estimate their fundamental parameters, using the technique described in \citet{Filippazzo2015}. Parameter values were determined using {\sc{SEDkit} V1.2.4}, which uses spectra, photometry and parallax measurements to create a distance-calibrated SED in order to determine the bolometric luminosity, $L_\mathrm{bol}$. This method uses photometric measurements to calibrate the spectral data. We use the same spectra detailed in Section \ref{sec:spectra}, and the photometry and parallaxes shown in Table \ref{tab:props}. Combining the $L_\mathrm{bol}$ value and age estimate, we use the \citet{Saumon2008} hybrid cloud evolutionary models to estimate the radius, mass, $\log$g and effective temperature ($T_{\mathrm{eff}}$) of each target. We present these determined values in Table \ref{tab:props}.

Our SED analysis highlights the similarities in the fundamental properties of \obj{s0136} and \obj{2m2139}, with $T_\mathrm{eff}$ within $\sim150~$K, masses that straddle the nominal boundary between planets and brown dwarfs, and almost identical radii. Investigating the atmospheres of this pair of twin worlds will offer insights on the diversity of extrasolar atmospheres.

\section{Retrieval Framework}

We perform atmospheric retrievals on our data using the \textit{\textit{Brewster}} framework \citep{Burningham2017, Burningham2021}.  We refer the reader to \citet{Burningham2017} for a complete description of the retrieval framework, but summarize the key features and new additions below.

\subsection{The Forward Model}
Our radiative transfer scheme computes the emergent flux from a layered atmosphere in the two-stream source approximation \citep{Toon1989} and includes scattering, as introduced by \citet{McKay1989} and adopted by numerous studies thereafter \citep[e.g.][]{Marley1996, Saumon2008,  Morley2012}. We use a 64 layer atmosphere  with geometric mean pressures in the range $\log P \mathrm{(bar)}=-4$ to $2.3$, spaced in $0.1$~dex intervals. The thermal profile is parameterized following \citet{Madhusudhan2009}. This scheme describes the atmospheres in three zones:

\begin{equation}
\begin{aligned}
    P_0<P<P_1:P=P_0 e ^{\alpha_1(T-T_0)^{\frac{1}{2}}}~~\mathrm{(Zone~1)} \\
    P_1<P<P_3:P=P_2 e ^{\alpha_2(T-T_2)^{\frac{1}{2}}}~~\mathrm{(Zone~2)}  \\
    P > P_3:T=T_3~~\mathrm{(Zone~3)}
\end{aligned}
\end{equation}
where $P_0$ and $T_0$ are the pressure and temperature at the top of the atmosphere, which becomes isothermal with temperature $T_3$ at pressure $P_3$. $P_0$ is fixed by the atmospheric model and parameters $T_0$ and $T_1$ can be eliminated due to continuity at the zonal boundaries. This leaves us with 6 free parameters: $\alpha_1$, $\alpha_2$, $P_1$, $P_2$, $P_3$, and $T_3$. Further ruling out a thermal inversion (i.e. setting $P2=P1$) simplifies this to 5 parameters for our thermal profile: $\alpha_1$, $\alpha_2$, $P_1$, $P_3$, and $T_3$.

\subsection{Gas Opacities}
We consider the following absorbing gases in our analysis: H$_2$O, CO, CO$_2$, CH$_4$, NH$_3$, CrH, FeH, SiO, Na and K. We chose to include these gases as they have been previously identified as important absorbing species in the spectra of L/T transition objects.

Layer optical depths due to absorbing gases are calculated using opacities sampled at $R=10,000$, and are sourced from \citet{Freedman2008, Freedman2014}, with updated opacities described in \citet{Burningham2017}. Line opacities are tabulated across our temperature-pressure range in $0.5$ dex steps in pressure and 20--500~K steps in temperature across the 75--4000~K range. This is linearly interpolated to our working pressure grid.

We include continuum opacities for H$_{2}$--H$_2$ and H$_2$--He collisionally induced absorption, using cross sections from \citet{Richard2012, Saumon2012}. We also include Rayleigh scattering due to H$_{2}$, He and CH$_{4}$, and continuum opacities due to bound-free and free-free absorption by H$^-$ \citep{John1988, Bell1987} and free-free absorption by H$^{-}_{2}$ \citep{Bell1980}. We use Na and K alkali opacities from \citet{Burrows2003}.


\subsection{Gas Phase Abundances}
We assume uniform-with-altitude mixing ratios for absorbing gases and constrain these abundances in the retrieval. This approach has been commonly used in retrievals \citep[e.g.][]{Burningham2017, Burningham2021, Gonzales2020}. While this simple assumption  cuts down on the number of parameters in the retrieval, it will not capture variations of gas abundances with altitude, which may be important. Thermochemical equilibrium models predict that abundances of many of the key absorbers are expected to vary by several orders of magnitude within the $0.1-10~$bar region from which we observe large contributions of flux \citep{Burningham2021}. 
In theory, retrievals of gas abundances that vary with altitude would be preferred, however the resulting large number of parameters produce a significant computational challenge. In an effort to address this issue, previous retrieval studies using the \textit{Brewster} framework have tested models that assume chemical equilibrium, drawing gas fractions for each layer of the atmosphere from thermochemical equilibrium grids. However, these models have not been preferred over the uniform-with-altitude mixing ratios  \citep{Burningham2017, Burningham2021, Gonzales2020} and we do not adopt these models in this work. 


\subsection{Cloud Model}\label{sec:cloudmodel}
Our cloud model is made up of three components -- 1) the location and structure of the cloud, 2) the optical properties of the cloud particles that define its opacity as a function of wavelength and 3) the existence of ``patches''.

\subsubsection{Location and structure of the cloud}
The vertical cloud structure can be described using two options -- a ``deck'' or ``slab'' cloud. Each structure is defined by how the cloud opacity varies with pressure. 
The ``deck'' cloud is the simpler of the two models, and is an optically-thick cloud where only the top of the cloud is visible. The deck cloud is parameterized by the pressure at which its total optical depth at $1~\mu$m passes unity, $P_\mathrm{deck}$, and a decay height, $\Delta \log P$, over which the optical depth decreases at shallower pressures and increases at deeper pressures as described in \citet{Burningham2021, Gonzales2020}. Deck clouds can become opaque rapidly with increasing pressure, such that minimal atmospheric information from below the deck cloud is accessible. This is important to keep in mind when interpreting retrieval results -- for atmospheric retrieval models containing a deck cloud, the profile below the deck cloud simply extends the gradient of the profile at the cloud deck.

The ``slab'' cloud differs from the deck in that this cloud is not forced to become optically thick at some pressure. Thus, along with the parameters that it shares with the deck cloud ($P_\mathrm{deck}$, $\Delta \log P$), the slab parameters also include the total optical depth at $1~\mu$m, $\tau_{\mathrm{cloud}}$. The optical depth is distributed through its extent as $d\tau/dP \propto P$, and reaches its total value at the high-pressure extent of the slab. We restrict the prior of the optical depth to the range $0-100$.

\subsubsection{Optical properties of the cloud particles}
We consider the wavelength dependent optical properties of different condensate species under the assumption of Mie scattering. We source optical data for condensates  from a variety of studies described in \citet{Burningham2021} and have pre-tabulated Mie coefficients as a function of radius and wavelength.
We calculate the wavelength optical depths, single scattering albedos and phase angles in each layer within the retrieval by integrating cross sections and Mie efficiencies over the particle size distribution in that layer for each condensate species. The total particle number density for a given condensate in each layer is calibrated to the optical depth at $1~\mu$m as calculated by the parameterized cloud model. In this work we only test retrieval models that assume a \citet{Hansen1971} distribution of particle sizes, since previous retrieval studies reported in \citet{Burningham2017} and \citet{Burningham2021} find that this particle size distribution provided better retrieved fits for cloudy brown dwarf spectra.

\subsubsection{Patchy cloud model}
In this paper we introduce a patchy cloud model to the \textit{Brewster} retrieval framework.
As our two targets are known to exhibit consistent, high-amplitude variability and are thus likely to have heterogeneous atmospheres, \obj{s0136} and \obj{2m2139} are the ideal targets to test this parametrization. 
Within \textit{Brewster}, patches are represented by individual atmospheric columns over which the radiative transfer is calculated. The flux from each column is then combined linearly according to their covering fractions to give the total emergent spectrum:
\begin{equation}
    F_{\mathrm{total}} = CF_{\mathrm(\mathrm{P}1)} + (1-C)F_{\mathrm(\mathrm{P}2)}
\end{equation}
where $F_{\mathrm{total}}$ is the total flux, $F_{\mathrm(\mathrm{P}1)}$ is the flux from Patch 1, $F_{\mathrm(\mathrm{P}2)}$ is the flux from Patch 2 and $C$ is the covering fraction. Within the \textit{Brewster} retrieval framework, each patch can contain $0-2$ different cloud structures. 
Only the cloud properties differ between columns -- the gas abundances and pressure-temperature profile remains the same between columns.

\subsection{The Retrieval Model}
The retrieval process consists of optimizing the parameters of the forward model such that the resultant spectrum provides the best match to the observed spectrum.
As described by \citet{Burningham2017}, we use a Bayesian framework to optimize the model fit to the data by varying the input parameters. 

Bayes' theorem provides a method to calculate the probability of a set of parameters $(x)$ given the data $(y)$, called the posterior probability:
\begin{equation}
    p(x|y) = \frac{\mathcal{L}(y|x)p(x)}{p(y)}
\end{equation}
where $\mathcal{L}$ is the likelihood of the data given the model parameters, $p(x)$ is the prior probability on the parameter set and $p(y)$ is the probability of the data marginalized over all parameter values. When using the above equation for parameter estimation, the denominator acts as a normalization factor, so to map out the posterior probability distribution $p(x|y)$, we only consider the two terms that make up the numerator.

We use the {\sc emcee} affine-invariant implementation of the Markov chain Monte Carlo method \citep{fm2013} to sample the posterior probability distribution. We use a log-likelihood function to assess the fit of the data to the model:
\begin{equation}
\ln \mathcal{L} (y|x) = -\frac{1}{2} \sum^n_{i=1} \frac{(y_i-F_i(x))^2}{s_i^2} - \frac{1}{2} \ln (2\pi s_i^2)
\end{equation}
where $y_i$ corresponds to the observed spectral flux points $i=0....n$, $s_i$ corresponds to the error on each flux point and $F_i$ are the forward model flux points for the current set of parameters, $x$. As described in \citet{Burningham2017} our errors are inflated using a tolerance parameter to allow for unaccounted sources of uncertainty.

\subsection{Model Selection}
We rank our models according to the Bayesian Information Criterion (BIC) in order of increasing $\Delta$BIC from the best-fit model.  The BIC is defined as
\begin{equation}
    \mathrm{BIC}=-2~\ln ~\mathcal{L}_\mathrm{max} + k ~\ln ~ N
\end{equation}
where $\mathcal{L}_\mathrm{max}$ is the maximum likelihood achievable by the model, $k$ is the number of parameters in the model and $N$ is the number of datapoints used in the fit \citep{Schwarz1978}. 

The model with the lowest BIC is the preferred model, and we follow the significance thresholds of \citet{Kass1995} for model preference:
\begin{itemize}
    \item $0<\Delta\mathrm{BIC}<2$: no preference worth mentioning
    \item $2<\Delta\mathrm{BIC}<6$: positive preference
    \item $6<\Delta\mathrm{BIC}<10$: strong preference
    \item $10<\Delta\mathrm{BIC}$: very strong preference
\end{itemize}

A variety of cloud parameterizations were explored for \obj{2m2139} and \obj{s0136}. Following the strategies from previous retrieval studies \citep{Burningham2017, Burningham2021}, we start with the simplest model available -- a cloud-free model -- and increase the complexity until we arrive at our best fit model for each target. Our model choices are driven by phase equilibrium chemistry and cloud modelling predictions of expected species \citep[e.g.][]{Burrows2001, Lodders2002Icar, Helling2008, Visscher2010} as well as  past detections of particular species in the literature \citep[e.g.][]{Cushing2006, Burningham2021, Luna2021}. 

\begin{table*}[tb]
\begin{tabular}{llll}
Model                  & $N$ Params           &\obj{s0136}   &  \obj{2m2139}\\     
                       &                      &$\Delta$~BIC  &$\Delta$~BIC\\ \hline    
Patchy Forsterite Slab \& Iron Deck & 32      & 0            & 0\\
Forsterite Slab \& Iron Deck  & 31            & 118          & 8\\
Enstatite Slab \& Iron Deck  & 31             & 119          &10\\
Patchy Enstatite Slab \& Iron Deck & 32       & 77           & 19\\
Forsterite Slab        & 27                   & 105          &20   \\
Iron Deck              & 26                   & 104          &38  \\
Enstatite Slab         & 27                   & 140          & 52\\
Forsterite Deck        & 26                   & 123          &56\\
No clouds              & 22                   & 578          & 56 \\ 
Enstatite Deck         & 26                   & 131          & 59     \\
Iron Slab              & 26                   & 120          & 65\\ \hline  \hline  
\end{tabular}
\caption{List of models tested in this work for  \obj{s0136} and \obj{2m2139} along with $\Delta$BIC. Models are arranged in order of the most favored model to least favored model for \obj{2m2139}.}
\label{tab:0136_2139_BICS}
\end{table*}


\begin{figure*}[tb]
   \centering
   \includegraphics[width=0.7\textwidth]{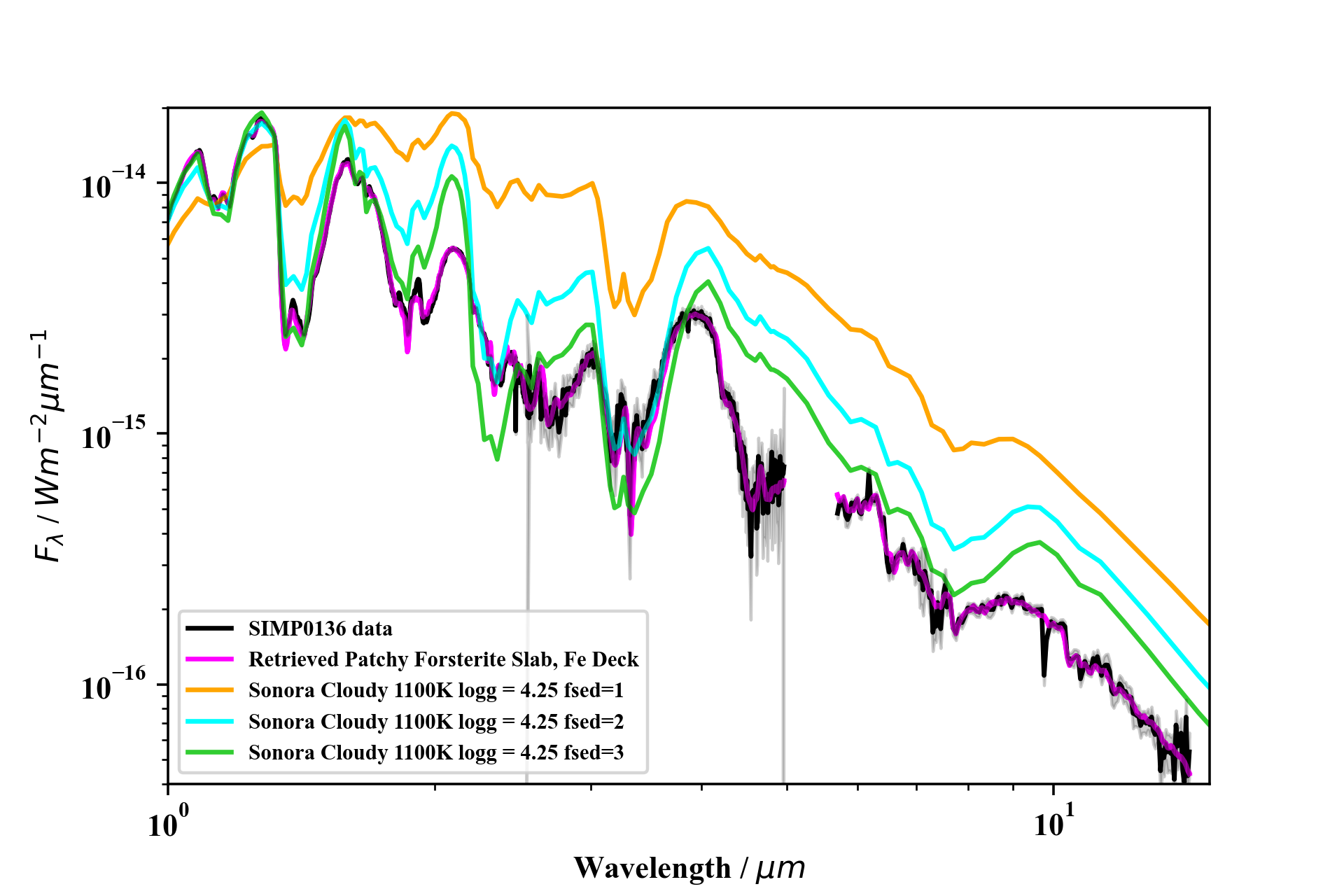}
   \caption{Maximum likelihood  retrieved model spectrum (pink) for the top-ranked model for \obj{s0136} overlaid with the data (black). Self-consistent grid models are shown for comparison, and are scaled to match the $J$-band flux in the observed spectrum.}
   \label{fig:0136_spectralfit}
\end{figure*}
 
\begin{figure*}[tb]
   \centering
   \includegraphics[width=0.7\textwidth]{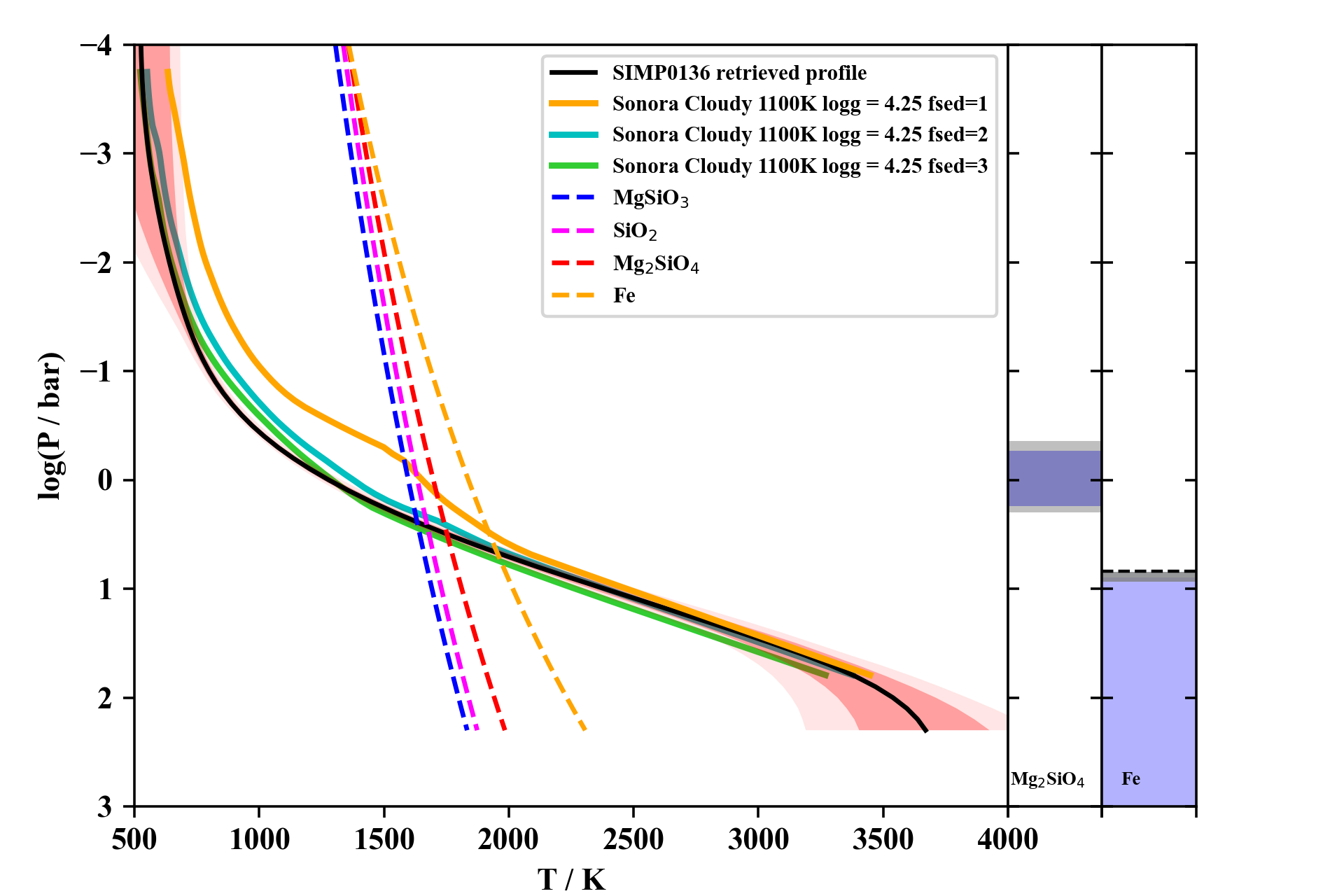}
   \caption{Retrieved thermal profile (black line, pink shading for 1$\sigma$ and 2$\sigma$ intervals) and cloud pressures for \obj{s0136}. Self-consistent  model profiles from the Sonora {Diamondback} (C. Morley et al. in prep) grid are plotted as solid colored lines. Phase-equilibrium condensation curves for possible condensate species are plotted as colored dashed lines. The cloud pressures for the forsterite (Mg$_2$SiO$_4$) slab and the iron (Fe) deck are indicated in bars to the left of the P/T profile. Purple shading indicates the median cloud location for the cloud, with grey shading indicating the 1$\sigma$ range. }
   \label{fig:0136_thermalprofile}
\end{figure*}

\section{SIMP~J0136+09  Retrieval Results}\label{sec:0136results}
We show the list of models tested for \obj{s0136} with the number of parameters and $\Delta$BIC for each model in Table \ref{tab:0136_2139_BICS}. 
The top-ranked model for \obj{s0136} is comprised of a patchy forsterite (Mg$_2$SiO$_4$) slab lying above a deep, iron (Fe) deck cloud. Out of the models we tested, the second best model has a $\Delta$BIC = 39, indicating that the patchy forsterite slab and iron deck model is very strongly preferred by the data.

Figure \ref{fig:0136_spectralfit} shows the model spectrum for the median set of parameters for the top-ranked model compared to the input data. The retrieved model reproduces the shape of the spectrum from $1-15~\mu$m. In Figure \ref{fig:0136_spectralfit} we also show self-consistent models from the Sonora {Diamondback} models (C. Morley et al. in prep) which are an extension of the Sonora grid models \citep{Marley2021,Karalidi2021} to cloudy atmospheres. The models shown have  $T_{\mathrm{eff}}=1100~$K, $\log g=4.25$ and sedimentation efficiency, $f_{\mathrm{sed}}=1-3$. The $f_{\mathrm{sed}}$ parameter, originating from the \citet{Ackerman2001} cloud model, controls the efficiency with which cloud particles settle out of the cloud.
Smaller $f_{\mathrm{sed}}$ values correspond to lower sedimentation efficiency, and thus vertically extended clouds composed of smaller particles while larger $f_{\mathrm{sed}}$ values correspond to efficient sedimentation which leads to physically thinner clouds composed of larger particles. 
The Sonora models capture the shape of the dominant absorbers but struggle to reproduce the overall shape of the spectrum. This is due to the lack of flexibility of the forward model compared to our retrieved model.

\subsection{Thermal Profile}
Figure \ref{fig:0136_thermalprofile} shows the retrieved thermal profile and cloud pressures for the winning  \obj{s0136}  model compared with self-consistent grid models and phase-equilibrium condensation curves and the Sonora Diamondback models. The models shown have  $T_{\mathrm{eff}}=1100~$K, $\log g=4.25$ and sedimentation efficiency, $f_{\mathrm{sed}}=1-3$. Models with $f_{\mathrm{sed}}=2-3$ {generally} agree with the retrieved profile at all pressures probed, {with the exception of the $\sim0.1-1~$bar region in which the model profiles are $>3\sigma$ hotter. This discrepancy highlights the advantages of using a flexible retrieval model to constrain atmospheric parameters and also the opportunity of using atmospheric retrieval results to inform future directions for forward models. The forward models solve for radiative-convective equilibrium given assumed elemental abundances and cloud properties. The physical processes underlying the departure of the retrieved profile from the forward model remains a topic of future investigation.}
The $f_{\mathrm{sed}}=1$ model provides a poorer match to the retrieved profile, but it is notable that the detached radiative zone in the model profile (visible as the kink in the pressure-temperature profile at $\sim1600~$K) approximately coincides with the pressure location of our retrieved forsterite slab cloud. Such detached radiative zones are driven by the formation of atmospheric clouds \citep{Tsuji2002,Burrows2006}, so their overlap in pressure space indicates that the retrieved cloud pressure level is consistent with the forward model.

\begin{deluxetable}{lrr}[tb]
\tabletypesize{\small}
\tablecolumns{10}
\tablewidth{0pt}
\setlength{\tabcolsep}{0.05in}
\tablecaption{Summary of retrieved gas abundances for the preferred models for \obj{s0136} and \obj{2m2139}.}
\label{tab:gases}
\tablehead{  
\colhead{}&
\colhead{\obj{s0136}}&
\colhead{\obj{2m2139}}}
\startdata
H$_2$O      & $-4.03^{+0.03}_{-0.03}$         & $-3.84^{+0.07}_{-0.06}$          \\
CO          & $-3.60^{+0.06}_{-0.06}$         & $-3.24^{+0.10}_{-0.09}$          \\
CO$_2$      & $-7.00^{+0.06}_{-0.07}$         & $-4.87^{+0.15}_{-0.74}$          \\
CH$_4$      & $-4.64^{+0.05}_{-0.04}$         & $-4.65^{+0.10}_{-0.09}$          \\
NH$_3$      & $-6.09^{+0.12}_{-0.13}$         & $-5.64^{+0.15}_{-0.17}$          \\
CrH         & $-9.75^{+0.12}_{-0.13}$         & $-9.47^{+0.16}_{-0.16}$          \\
FeH         & $-9.33^{+0.06}_{-0.06}$         & $-9.68^{+0.14}_{-0.18}$          \\
SiO         & $-9.09^{+1.95}_{-1.97}$         & $-9.55^{+2.02}_{-1.59}$          \\
Na+K        & $-6.39^{+0.08}_{-0.08}$         & $-6.51^{+0.15}_{-0.15}$          \\
log(g)      & $4.25^{+0.08}_{-0.08}$          & $4.27^{+0.16}_{-0.13}$        
\enddata
\end{deluxetable}
\begin{deluxetable*}{lcccc}[tb]
\tabletypesize{\small}
\tablecolumns{10}
\tablewidth{0pt}
\setlength{\tabcolsep}{0.05in}
\tablecaption{Summary of retrieved cloud properties for the preferred models for \obj{s0136} and \obj{2m2139}.}
\label{tab:cloudprops}
\tablehead{  
\colhead{}&
\colhead{\obj{s0136}}&
\colhead{ }&
\colhead{\obj{2m2139}}&
\colhead{ }}
\startdata
Cloud No.                               & 1                          & 2                          &  1                        &  2 \\
Type                                    & Slab                       & Deck                       & Slab                      & Deck                  \\
Species                                 & Mg$_2$SiO$_4$                 & Fe                      & Mg$_2$SiO$_4$                 & Fe                    \\
Cloud Coverage                          & $0.70^{+0.03}_{-0.04}$     &n/a                         &$0.83_{-0.06}^{+0.06}$        &n/a                       \\
Max $\tau_{\mathrm{cloud}}$ at $1~\mu$m & $12.03^{+1.03}_{-0.85}$    &n/a                         & $8.17_{-1.38}^{+1.54}$     &n/a                    \\
Reference Pressure / log$P$ (bar) & $0.24^{+0.06}_{-0.05}$ (max $\tau$) & $0.90^{+0.04}_{-0.05}$ ($\tau=1$)  & $0.12_{-0.09}^{+0.08}$ (max $\tau$)& $0.86_{-0.07}^{+0.09}$ ($\tau=1$)    \\
Height / $d$log$P$(bar)                      & $0.50^{+0.09}_{-0.096}$    & $0.06^{+0.03}_{-0.02}$     & $0.32_{-0.15}^{+0.20}$     & $0.03_{-0.02}^{+0.03}$     \\
Log(effective particle radius $a$/$\mu$m)        & $-1.47^{+0.14}_{-0.14}$  & $-1.35^{+0.25}_{-0.31}$    & $-1.21_{-0.19}^{+0.18}$    & $-1.31_{-0.66}^{+0.45}$     \\
Particle radius spread (Hansen distribution $b$) & $0.36^{+0.33}_{-0.25}$     & $0.51^{+0.33}_{-0.33}$     & $0.42_{-0.29}^{+0.34}$     & $0.52_{-0.35}^{+0.34}
$    
\enddata
\end{deluxetable*}

\subsection{Cloud Properties}\label{sec:0136_cloudprops}
The preferred model is comprised of a patchy forsterite slab cloud and a deeper iron deck cloud. A summary of the retrieved parameters for the preferred models is presented in Table \ref{tab:cloudprops}. The top ranked model is a relatively complicated one that consists of two types of clouds, one of which is patchy. This model complexity is enabled by our large wavelength range, as was also found by \citet{Burningham2021} who found a similarly complex best-fit model for \obj{2m2224} using this wavelength range.

The median pressures of these clouds are indicated by the shaded regions in the right-hand panels of Figure \ref{fig:0136_thermalprofile}. 
The forsterite slab cloud is the highest altitude cloud in the model, with a cloud base pressure of $\log P (\mathrm{bar}) = 0.24^{+0.06}_{-0.05}~$dex. Both its height and depth are well-constrained by the retrieval. The iron deck is located deeper than the forsterite slab, with a well-constrained $\tau_\mathrm{Fe}=1$ pressure level of $\log P (\mathrm{bar}) = 0.90^{+0.04}_{-0.05}~$dex. The pressure change over which the optical depth of the iron clouds drops from 1.0 to 0.5 is constrained to $d$log$P=0.06^{+0.03}_{-0.02}$. This represents a cloud with a small decay height, that is confined to a narrow range of pressures. This cloud is notably different to the iron deck retrieved for the L dwarf 2M2224-01 by \citet{Burningham2021}, whose iron deck cloud has a similar $\tau_{\mathrm{Fe}}=1$ pressure level, but with a much larger scale height. This is likely driven by the lower temperatures of the T2 type \obj{s0136}, which results in the formation of iron clouds at deeper pressures.

Figure \ref{fig:0136_thermalprofile} also shows the condensation curves for several species, calculated for solar composition gas following \citet{Visscher2010}. Condensation curves provide a useful guide on the species that can condense in the atmosphere, but do not indicate whether that particular condensate will necessarily form. The pressure depths of both retrieved clouds are consistent with being located to the left of their respective condensation curves on the retrieved thermal profile -- they are both located at cooler temperatures and shallower pressures than the condensation point and are thus consistent with condensation chemistry predictions.

As discussed in Section \ref{sec:cloudmodel}, our models use Mie scattering clouds with a Hansen distribution for particle sizes.
For \obj{s0136} we find that both cloud species are dominated by sub-micron sized grains with a negligible number of particles larger than $1~\mu$m.


Our top ranked model is one that contains two ``patches'', as discussed in Section \ref{sec:cloudmodel}. Patch 1 consists of the forsterite slab cloud and an iron deck cloud, while Patch 2 consists of the deeper iron deck cloud. This parameterization adds only one additional parameter to the model -- the covering fraction of Patch 1. Our retrieval constrains the coverage  of Patch 1 to $70^{+3}_{-4}\%$ of the surface. The top ranking of a cloud model consisting of multiple patches supports the idea of an inhomogeneous atmosphere, as first concluded from time-resolved variability monitoring \citep{Artigau2009, Apai2013}. We will discuss the implications of this patchy model in Section \ref{sec:patchyclouds}.

\subsection{Bulk Properties}\label{sec:0136_bulkprops}

Figure \ref{fig:0136_gascorner} shows the posterior distributions for the retrieved gas fractions and log$~g$,  along with derived values for radius, mass, atmospheric metallicity and C/O ratio, and extrapolated $T_{\mathrm{eff}}$. We also present these values in Table \ref{tab:props}. The radius is estimated using the retrieved model scaling factor and Gaia parallax. The mass is then computed using the derived radius and the retrieved log$~g$. $L_{\mathrm{bol}}$ is found by extrapolating the retrieved model to cover the $0.5-20~\mu$m range, summing the flux and scaling by $4 \pi D^2$, where $D$ is the distance. $T_{\mathrm{eff}}$ is then determined using the extrapolated $L_{\mathrm{bol}}$ and inferred radius. The atmospheric C/O ratio is estimated under the assumption that all carbon and oxygen in the atmosphere exist within the absorbing gases included in the retrieval. The metallicity is estimated by considering elements within our retrieved absorbing gases and comparing their inferred abundances to their solar values \citep{Asplund2009}.

We find that some of the fundamental parameters derived from our retrieval show poor consistency with the SED-derived values.  We discuss these discrepancies in more detail in Section \ref{sec:SED_comparison}.

\section{2MASS~J2139+02   Retrieval Results}
Table \ref{tab:0136_2139_BICS} shows the list of models tested for \obj{2m2139} along with the number of parameters for each model and the resulting $\Delta$BIC. The winning model for \obj{2m2139} consists of a patchy forsterite slab cloud and an iron deck cloud -- the very same model that is preferred for our twin object \obj{s0136}. With a $\Delta$BIC$=8$ between the winning and second best model, the patchy  forsterite slab and iron deck model is very strongly preferred over all of the models tested.

\begin{figure*}[tb]
   \centering
   \includegraphics[width=0.7\textwidth]{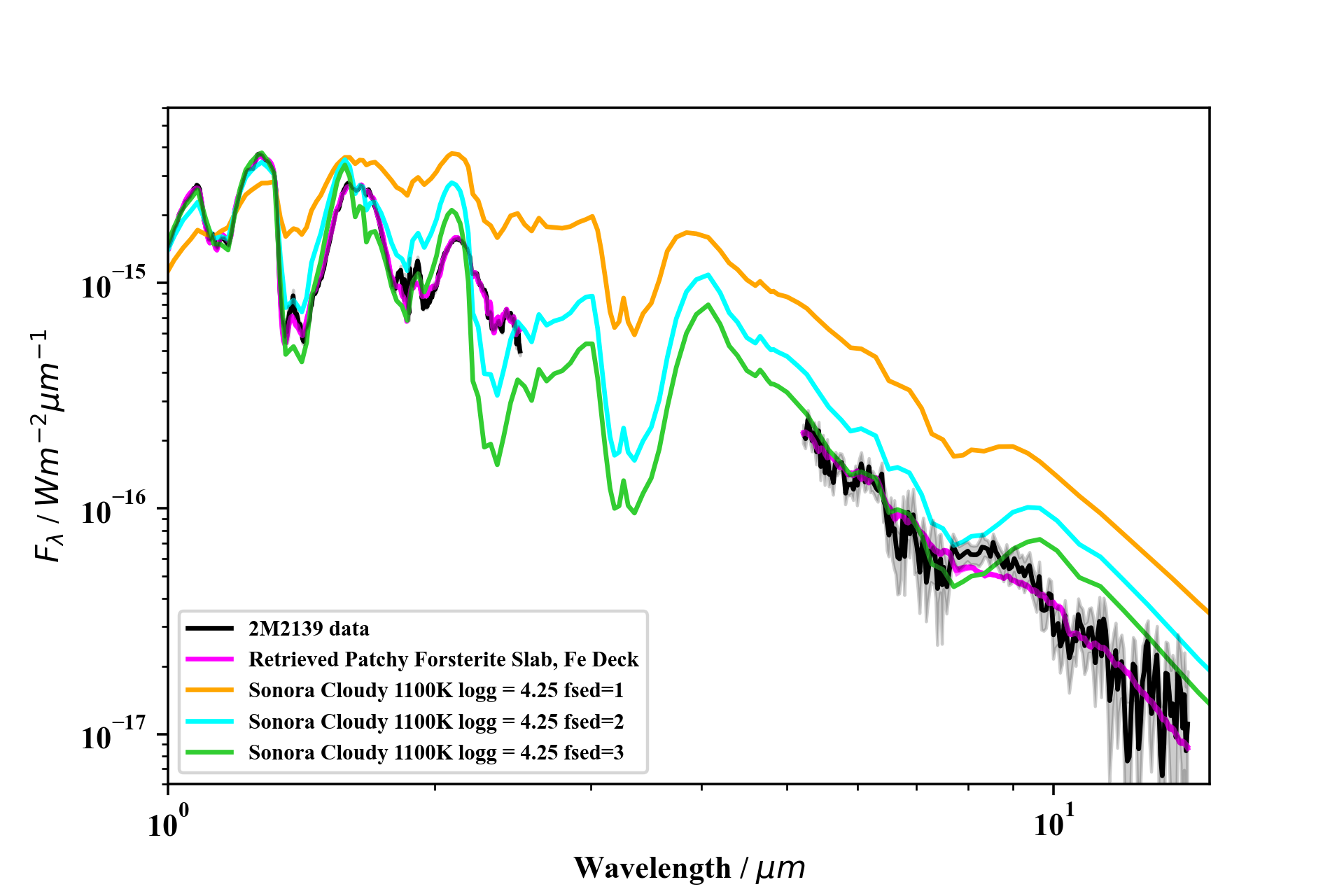}
   \caption{Maximum likelihood  retrieved model spectrum (pink) for the top-ranked model for \obj{2m2139} overlaid with the data (black). Self-consistent grid models are shown for comparison, and are scaled to match the $J$-band flux in the observed spectrum.}
   \label{fig:2139_spectral_fit}
\end{figure*}

In Figure \ref{fig:2139_spectral_fit} we show the top ranked model spectral fit along with the observed spectrum. The retrieved model spectrum fits the entire $1-15~\mu$m spectral range very well. We also show the Sonora {Diamondback} models in Figure \ref{fig:2139_spectral_fit}. Similarly to the spectral fit presented for \obj{s0136} in Section \ref{sec:0136results}, the retrieval provides a better fit to the overall spectrum due to the flexibility of the model. 


\begin{figure*}[tb]
   \centering
   \includegraphics[width=0.7\textwidth]{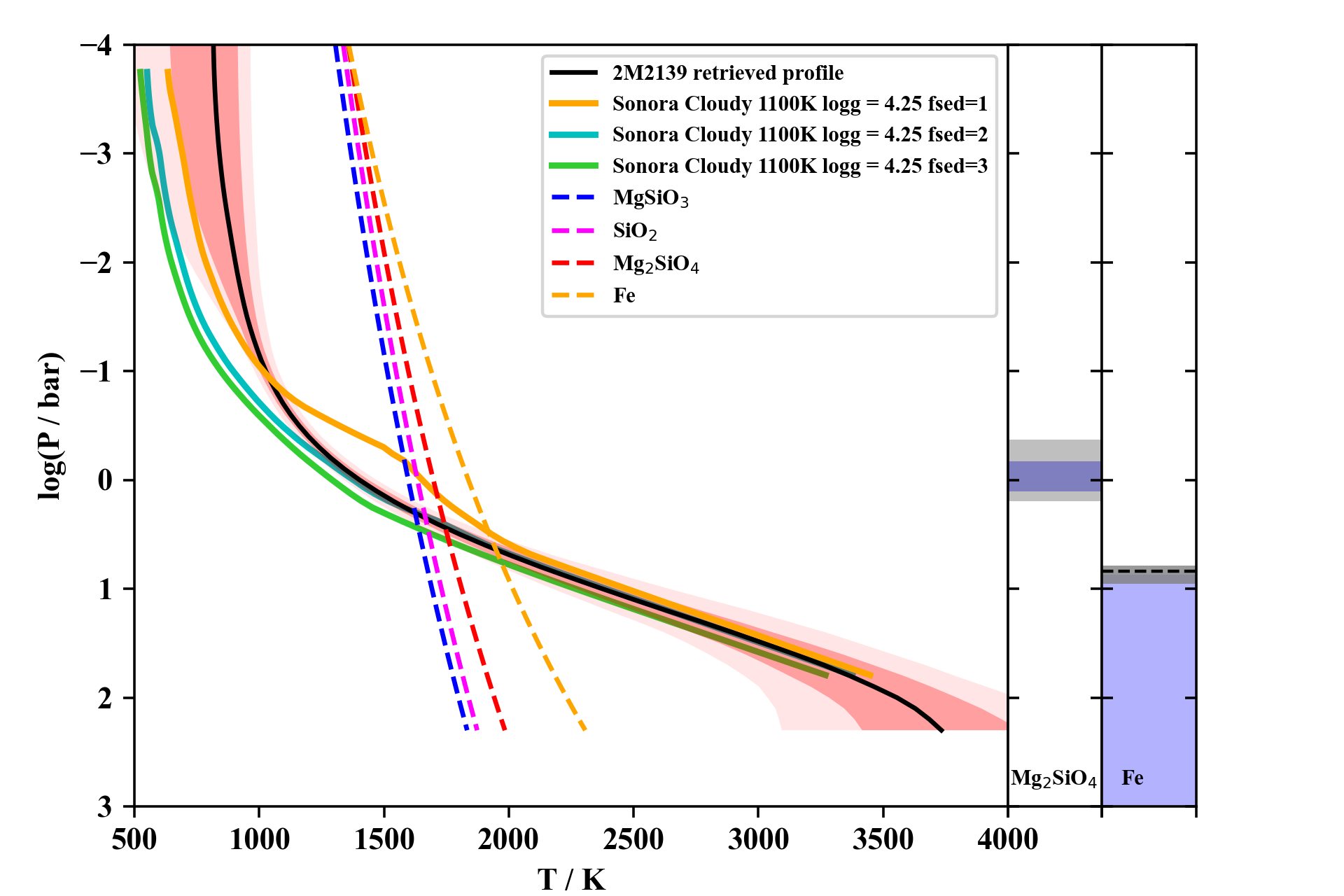}
   \caption{Retrieved thermal profile (black line, pink shading for 1$\sigma$ and 2$\sigma$ intervals) and cloud pressures for \obj{2m2139}. Self-consistent  model profiles from the Sonora {Diamondback} (C. Morley et al. in prep) grid are plotted as solid colored lines. Phase-equilibrium condensation curves for possible condensate species are plotted as colored dashed lines. The cloud pressures for the forsterite (Mg$_2$SiO$_4$) slab and the iron (Fe) deck are indicated in bars to the left of the P/T profile. Purple shading indicates the median cloud location for the cloud, with grey shading indicating the 1$\sigma$ range.}
   \label{fig:2139_thermalprofile}
\end{figure*}

\subsection{Thermal Profile}

Figure \ref{fig:2139_thermalprofile} shows the retrieved thermal profile and cloud pressure for the winning model compared to self-consistent grid models and phase-equilibrium condensation curves. As in Figure \ref{fig:0136_thermalprofile}, we show Sonora {Diamondback} models as comparison (C. Morley et al. in prep), and show the  $T_{\mathrm{eff}}=1100~$K, log$g=4.25$ and sedimentation efficiency, $f_{\mathrm{sed}}=1-3$. The $f_{\mathrm{sed}}=2-3$ models show good agreement with our retrieved thermal profile below 0.3 bar, however our retrieved profile is slightly warmer than both models at shallower pressures.

\subsection{Cloud Properties}
The spectrum of \obj{2m2139} is best described by a patchy forsterite slab cloud and an iron deck cloud. We present a summary of the cloud parameters for \obj{2m2139} in Table \ref{tab:cloudprops}.

The pressures of the two cloud species in the atmosphere of \obj{2m2139} are shown by the shaded regions to the right of Figure \ref{fig:2139_thermalprofile}. The forsterite slab lies at a shallower pressure in the atmosphere, with a well-constrained base pressure of $\log P=0.12_{-0.09}^{+0.08}$~dex and a height of $d\log P=0.32_{-0.15}^{+0.20}$ dex. The coverage of the patchy forsterite slab cloud is constrained to $83\pm6\%$ of the surface. These parameters are very similar to those retrieved for \obj{s0136}.

The iron cloud deck becomes optically thick at deeper pressures of $\sim7$~bar. The location of the $\tau_{\mathrm{Fe}}=1.0$ pressure level is tightly constrained to $\log P$~(bar)$=0.86_{-0.07}^{+0.09}$~dex. The median value for the pressure over which the optical depth drops from $\tau_{\mathrm{Fe}}=1.0$ to $\tau_{\mathrm{Fe}}=0.5$ is small ($d log P=0.03$), which indicates a compact cloud that is confined to deeper pressures.

Figure \ref{fig:2139_thermalprofile} also shows the condensation curves for several species of interest, as discussed in Section \ref{sec:0136_cloudprops}. The forsterite slab cloud is placed at pressures and temperatures consistent with those expected from forsterite condensation curves. The $\tau_{\mathrm{Fe}}=1.0$ pressure level for the iron deck cloud lies approximately at the intersection of the P/T profile and the Fe condensation curve and its pressure location is thus consistent with condensation curve chemistry.

All of the models we tested used Mie scattering and Hansen particle distributions. For \obj{2m2139}, the slab cloud is dominated by sub-micron sized particles (median size $\sim0.06~\mu$m), with a negligible number of particles larger than $1~\mu$m. The deck cloud is also composed of sub-micron particles (median size $\sim0.05~\mu$m).

\subsection{Bulk Properties}
Figure \ref{fig:2139_gascorner} shows the posterior distributions for the retrieved gas fractions for absorbing gases and $\log g$ for our winning model for \obj{2m2139}. We derive values for radius, mass, atmospheric metallicity, C/O ratio and $T_{\mathrm{eff}}$ using the method discussed in Section \ref{sec:0136_bulkprops}. These values are also given in Table \ref{tab:props}.

As in the case of \obj{s0136}, we find that the retrieved and SED-derived values are generally inconsistent with each other, with the exception of $L_\mathrm{bol}$ and $\log g$. These discrepancies are discussed in detail in Section \ref{sec:SED_comparison}.


%

\subsection{Discrepant Feature in the Spectrum of 2MASS~J2139+02}

While the retrieved spectral fit for \obj{2m2139} (Figure \ref{fig:2139_spectral_fit}) provides a good overall fit, it is clear that the best-fit retrieval fails to adequately fit the $7-10~\mu$m region. In fact, none of the models tested on this spectrum provided a good fit in this part of the spectrum -- all models underestimate the flux in this region.

Our first assumption was that this may be due to an overestimation of the abundance of one of our absorbing gases within the model. This could arise due to our assumed vertically constant abundances whereby a good fit in one part of the spectrum may force a poor fit in another part of the spectrum if the abundance of an absorbing gas changes rapidly through the atmosphere. 
However, having compared the spectral contributions of each of our absorbing gases, we find that none of these gases uniquely coincide with this region. We conclude that the discrepancies between the observed and retrieved spectra are not caused by an over-estimation of the abundances of the gases within our model.


We also consider whether systematics may cause the feature in the spectrum. In particular, the overlap region between the SL2 and SL1 modes ($7.4-8.7~\mu$m) approximately corresponds to the region of discrepancy between data and model. To combine the spectra from each mode, \citet{Suarez2022} use SL1 spectra at wavelengths $>7.5~\mu$m. Having obtained both SL1 and SL2 mode spectra (G. Suarez, priv. comm.) and compared the spectra in the overlap region, we find that the spectra agree across modes, indicating that this feature is likely not caused by a stitching error.

\begin{table}
\begin{tabular}{lll}
Model                  & $N$ Params           &$\Delta$~BIC  \\ \hline    
No clouds              & 20                   & 280           \\
Forsterite Slab \& Iron Deck  & 29           & 8      \\
Patchy Forsterite Slab \& Iron Deck& 32           & 0      \\ 
Patchy Enstatite Slab \& Iron Deck & 32           & 42      \\  \hline  \hline
\end{tabular}
\caption{Models tested in this work for the \obj{2m2139} spectrum without the $7-10~\mu$m region with their corresponding $\Delta$BIC values. The patchy forsterite and iron deck model is still the highest ranking model when this wavelength region is removed.}
\label{tab:2139_cutout}
\end{table}

For the purposes of this work, the most important question is whether this region is driving our model ranking or the parameters within our best fit model. To answer this, we performed a smaller set of retrievals on a spectrum of \obj{2m2139} with the $7-10~\mu$m region removed. We show the resulting $\Delta$BIC values for these runs in Table \ref{tab:2139_cutout}. The patchy forsterite and iron deck model is still the highest ranking model when this wavelength region is removed, and the retrieved parameters are consistent across both models.


\section{Discussion}

\subsection{Comparison with SED-Derived Fundamental Parameters}\label{sec:SED_comparison}


We find consistency between our extrapolated and retrieved $L_\mathrm{bol}$ and $\log g$ values and the SED-derived values (shown in Table \ref{tab:props}). In both cases the retrieval $L_\mathrm{bol}$ value is slightly higher than the SED value, but falls within the range expected from past retrieval studies \citep{Gonzales2020, Gonzales2022, Calamari2022}. We also find that our retrieved and SED-derived $\log g$ values are consistent.

Our derived radii for \obj{2m2139} and \obj{s0136} are considerably smaller than evolutionary models predictions. 
Relatedly, our high derived $T_\mathrm{eff}$ values are driven by the small retrieved radius. The low derived masses are also likely caused by the small retrieved radius. The retrieved $\log g$ is driven by the impact of the scale height of the atmosphere on spectral features, and is found to be consistent with evolutionary model predictions. However, when combined with the small radius, it implies a very low mass that is inconsistent with evolutionary model predictions.

There are many instances in the literature of studies that retrieve unexpectedly small radii, particularly in the mid L to early T spectral type regime. Using the \textit{Brewster} retrieval framework, \citet{Burningham2021} found a similarly small radius for the cloudy L dwarf \obj{2m2224}, the only other object retrieved with a similar wavelength range. Small radii have also been found for L and L/T transition dwarfs using other retrieval frameworks \citep[e.g.][]{Kitzmann2020, Molliere2020, Lueber2022, ZhangSnellen2021} and self-consistent modelling frameworks \citep[e.g.][]{Barman2011, Sorahana2013, Brock2021}. Additionally, \citet{Zalesky2019} report anomalously small radii for a sample of late-T and early Y dwarfs. In contrast, the retrieved radii for mid type T dwarfs are generally consistent with evolutionary models \citep[e.g.][]{Calamari2022, Gonzales2020, Line2017}. The prevalence of this problem for objects with temperatures where clouds are thought to be abundant in the atmosphere suggests that the problem may be linked to the existence of clouds. As discussed by \citet{Burningham2021}, the existence of a grey cloud at very shallow pressures would act to reduce the total flux and thus allow the radius to increase. However, such a cloud would require much larger particle sizes than those that have been retrieved in this or other retrieval studies in order to produce a primarily grey opacity.

\subsection{Patchy Forsterite Clouds for Two Variable Planetary-Mass Objects} \label{sec:patchyclouds}
The winning models for \obj{s0136} and \obj{2m2139} are strikingly similar, the preferred models consists of a patchy forsterite slab cloud above an iron deck in each case. Since our parameterization of cloud opacity does not incorporate cloud condensation models to determine which clouds are favored, it is encouraging that the retrieved presence of forsterite is also predicted by a range of microphysical and phase equilibrium models at solar composition \citep[e.g.][]{Helling2006,LoddersFegley2006, Visscher2010, Gao2020}. {While this model is very strongly preferred by the data for both targets, the $\Delta$BIC values are more decisive for \obj{s0136} across all models (see Table \ref{tab:0136_2139_BICS}). This is primarily due to the inclusion of the AKARI/IRC spectrum for \obj{s0136} which provides more data points to assess each model, and secondarily due to the higher SNR of the spectra for \obj{s0136} (see Section \ref{sec:spectra} for a discussion of the spectra and their SNR).}

\begin{figure*}[tb]
   \centering
       \includegraphics[width=0.49\textwidth]{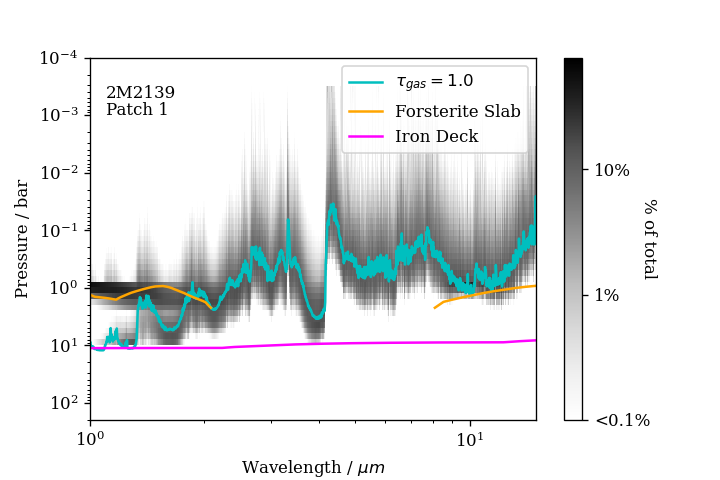}
   \includegraphics[width=0.49\textwidth]{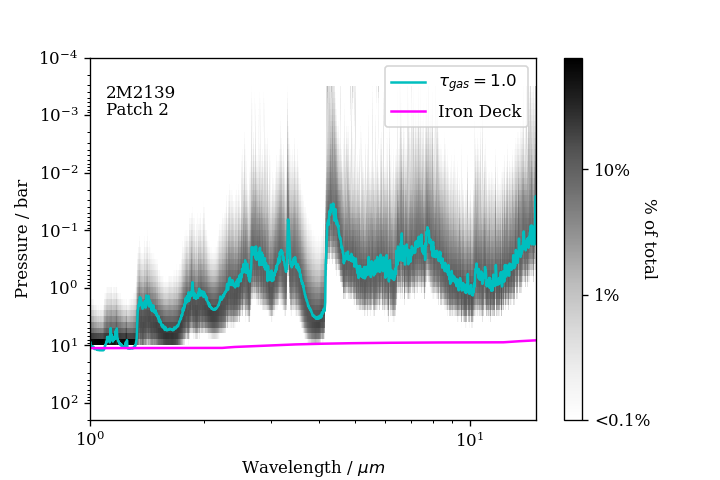}\\
      \includegraphics[width=0.49\textwidth]{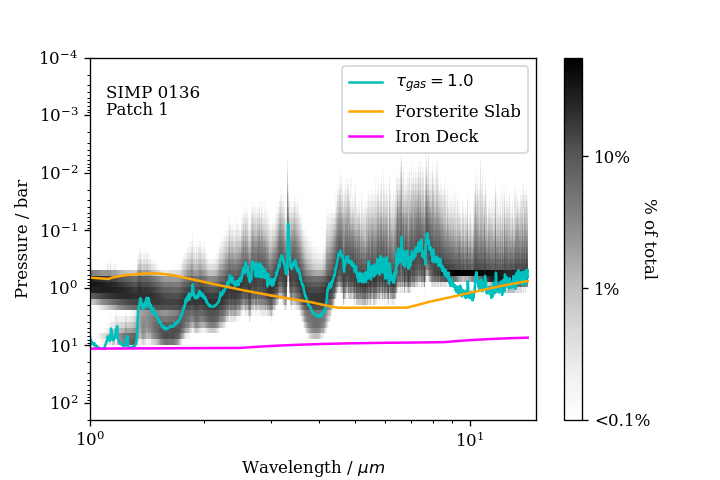}
       \includegraphics[width=0.49\textwidth]{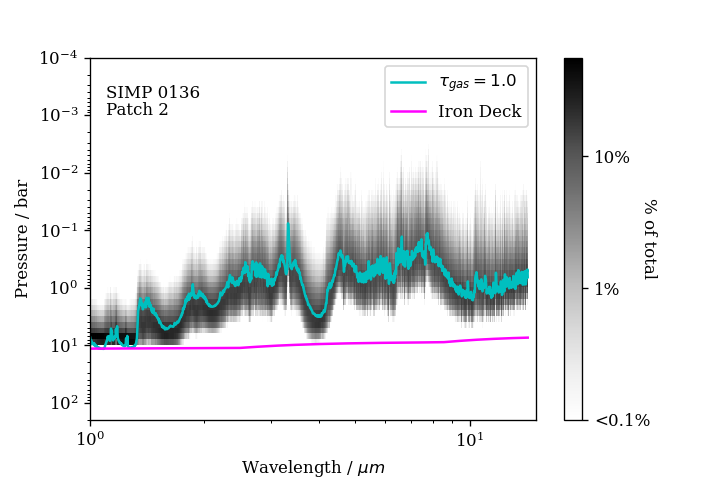}
   \caption{Contribution functions for \obj{2m2139} (top) and \obj{s0136} (bottom) based on the maximum likelihood retrieved parameters for the top-ranked model in each case. We show separate contribution functions for Patch 1 (forsterite slab above an iron deck, left) and Patch 2 (iron deck, right). The black shading shows the percentage contribution of each pressure to the flux at each wavelength. $\tau=1$ lines are included for gas phase opacities (cyan), the forsterite slab (orange) and the iron deck (pink). For \obj{2m2139}, the forsterite slab cloud does not become optically thick at wavelengths of $\sim2-8~\mu$m, so there is a break in this line.}
   \label{fig:contribution_functions}
\end{figure*}

The results presented here demonstrate for the first time that a patchy model has been preferred within a retrieval framework. 
The fact that both targets are photometrically and spectroscopically variable \citep[e.g.][]{Artigau2009, Radigan2012, Apai2013} provides independent evidence for the heterogeneous nature of their atmospheres, and was thus one of our main motivators for testing a patchy cloud model to reproduce their spectra.


Beyond the fact that the same model is preferred for both of our targets, the properties of the retrieved clouds are very similar for both \obj{s0136} and \obj{2m2139}. We show the cloud properties for the preferred model for each object in Table \ref{tab:cloudprops}. In both cases, the patchy forsterite cloud has a similar level of coverage (70\% for \obj{s0136} and 83\% for \obj{2m2139}). The forsterite clouds are situated at similar pressures of $1.3-1.7$~bar, and have similar heights ($2-5$~bar). The optical depth of \obj{s0136}'s forsterite cloud is marginally larger. The forsterite slab for both objects is comprised of sub-micron particles of radius $0.03-0.06~\mu$m with a negligible number of particles with radii above $1~\mu$m. Similarly, the properties of the iron deck cloud are consistent across both targets -- the deck cloud in both cases is situated at $7-8~$bar and is composed of sub-micron sizes grains.

We show the contribution functions for each model in Figure \ref{fig:contribution_functions}. The contribution function in an atmospheric layer that lies between pressures $P_1$ and $P_2$ is defined as 
\begin{equation}
    C(\lambda, P) = \frac{B(\lambda,T(P))\int_{P_1}^{P2}d\tau}{\mathrm{exp}\int_0^{P_2}d\tau}
\end{equation}
where $B(\lambda,T(P))$ is the Planck function. Since a patchy model is preferred for both \obj{2m2139} and \obj{s0136}, we obtain two contribution functions for each object. Patch 1 consists of the forsterite slab and an iron deck while Patch 2 consists of solely the iron deck cloud. As we expect from the similarity between the retrieved parameters for both targets, the contribution functions for \obj{2m2139} and \obj{s0136} are remarkably similar. In Patch 1, we see that the forsterite slab contributes to the flux at $\sim1-2~\mu$m, $\sim4~\mu$m and at $\sim9-11~\mu$m for both models, with the gas opacity contributing flux in other wavelength regions. In Patch 2, we note a contribution of the iron deck in a small wavelength range shortward of $\sim1.5~\mu$m, with the gas opacity dominating elsewhere.

\subsection{Linking the Observed Variability Signatures to Brewster Retrieval Results}

A major goal of this work is to link our retrieval results with the observed spectral variability signatures previously reported in the literature for our targets, specifically the HST/WFC3 monitoring observation presented by \citet{Apai2013} (and discussed in detail in Section \ref{sec:targets}). Based on the observed spectral variability, the authors used the color information to provide a picture of the atmospheric cloud structures responsible for the observed variability. They conclude that the spectral behavior of both targets are best described by a heterogeneous mixture of high-altitude, cool, thick clouds and deeper, hotter thin clouds. In this section we investigate whether the observed variability could be explained using our top ranked model. 

\begin{figure}
   \centering
       \includegraphics[width=0.5\textwidth]{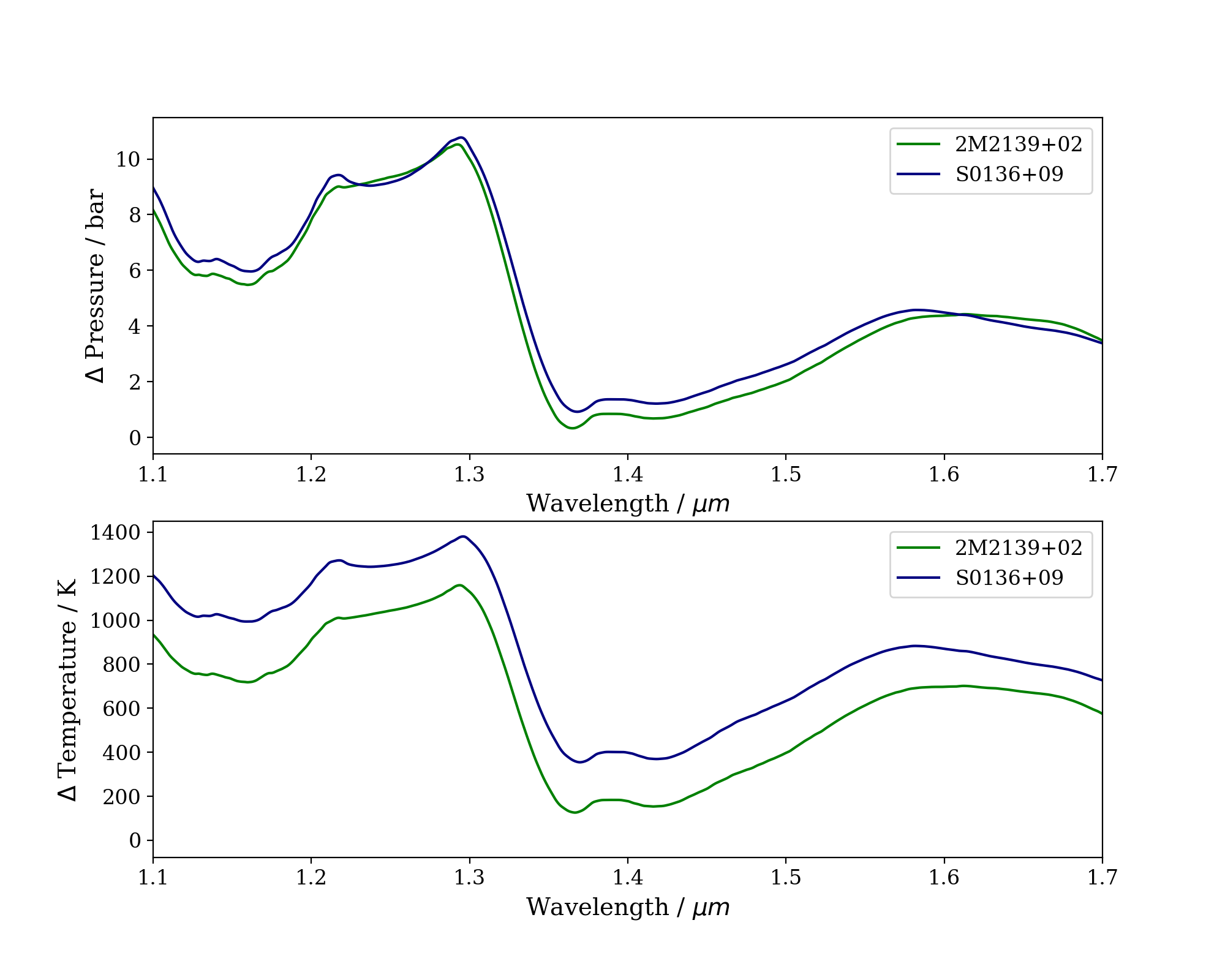}
   \caption{The change in pressure, $\Delta P$, (top) and the change in temperature, $\Delta T$, (bottom) for \obj{2m2139} (blue) and \obj{s0136} (purple) between Patch 1 (iron deck and forsterite slab) and Patch 2 (iron deck only). We determine the photosphere of each patch as the pressure at which we reach $\tau=1$ due to either cloud or gas opacity. When then calculate $\Delta P$ as the difference in pressures and  $\Delta T$ as the difference in temperatures between the photospheres of the two patches of our model. $\Delta P$ and $\Delta T$ are thus a measurement of how much deeper and hotter we observe into the atmosphere in Patch 2 compared to Patch 1. The high $\Delta T$ values suggest that much of the coverage for Patch 1 is distributed homogeneously. The shape of both   $\Delta T$ and  $\Delta P$ as a function of wavelength broadly match the shape of the observed variability reported by \citet{Apai2013}.}
   \label{fig:variability_pressure_temperature}
\end{figure}

Using the contribution functions presented in Figure \ref{fig:contribution_functions}, we can begin to approximate the variability that may be caused by the patchy forsterite slab cloud. 
We determine the photosphere of each patch as the pressure at which we reach $\tau=1$ due to either cloud or gas opacity. We then calculate $\Delta P$ as the difference in pressures and  $\Delta T$ as the difference in temperatures of the photosphere between the two patches. $\Delta P$ and $\Delta T$ are thus a measurement of how much deeper and hotter we observe into the atmosphere in Patch 2 (consisting of the iron deck cloud only) compared to Patch 1 (consisting of the forsterite slab and the iron deck clouds). 
In Figure \ref{fig:variability_pressure_temperature}, we show the change in pressure (top) and temperature (bottom) between Patch 1 and Patch 2 for both targets. Since we find very similar cloud structures for \obj{s0136} and \obj{2m2139}, it is not surprising that the changes in pressure and temperature are very similar. The temperature changes are larger for \obj{s0136} due to a slightly steeper slope in the P/T profile at the pressures between cloud layers (shown in Figure \ref{fig:0136_thermalprofile}). 

The large $\Delta T$ values between the photosphere of each patch are notable. If the variability were driven by a single region defined by Patch 1, with a temperature that is $200-1200~$K cooler than the second region defined by Patch 2, one would expect an amplitude much larger than that observed for either object. We argue that these results imply that the regions that define Patch 1 are distributed mostly homogeneous as opposed to forming one large region in the atmosphere. To achieve the variability amplitudes that have been measured for \obj{s0136} and \obj{2m2139}, the heterogeneity in silicate cloud clover must represent a small fraction of the total silicate cloud covering fractions that we have retrieved for our targets. We investigate this idea in more detail in the following analyses. 

We also find that the spectral behavior of $\Delta P$ and $\Delta T$ broadly match the shape of the observed variability: the highest variability amplitude occurs at $\sim1.25~\mu$m, where $\Delta P$ and $\Delta T$ are the highest, while the lowest variability amplitude is observed in the water band at $\sim1.4~\mu$m where $\Delta P$ and $\Delta T$ are at their lowest. This broad agreement in shape suggests that the variability may be driven by the patchy forsterite cloud.

\begin{figure*}
   \centering
       \includegraphics[width=0.75\textwidth]{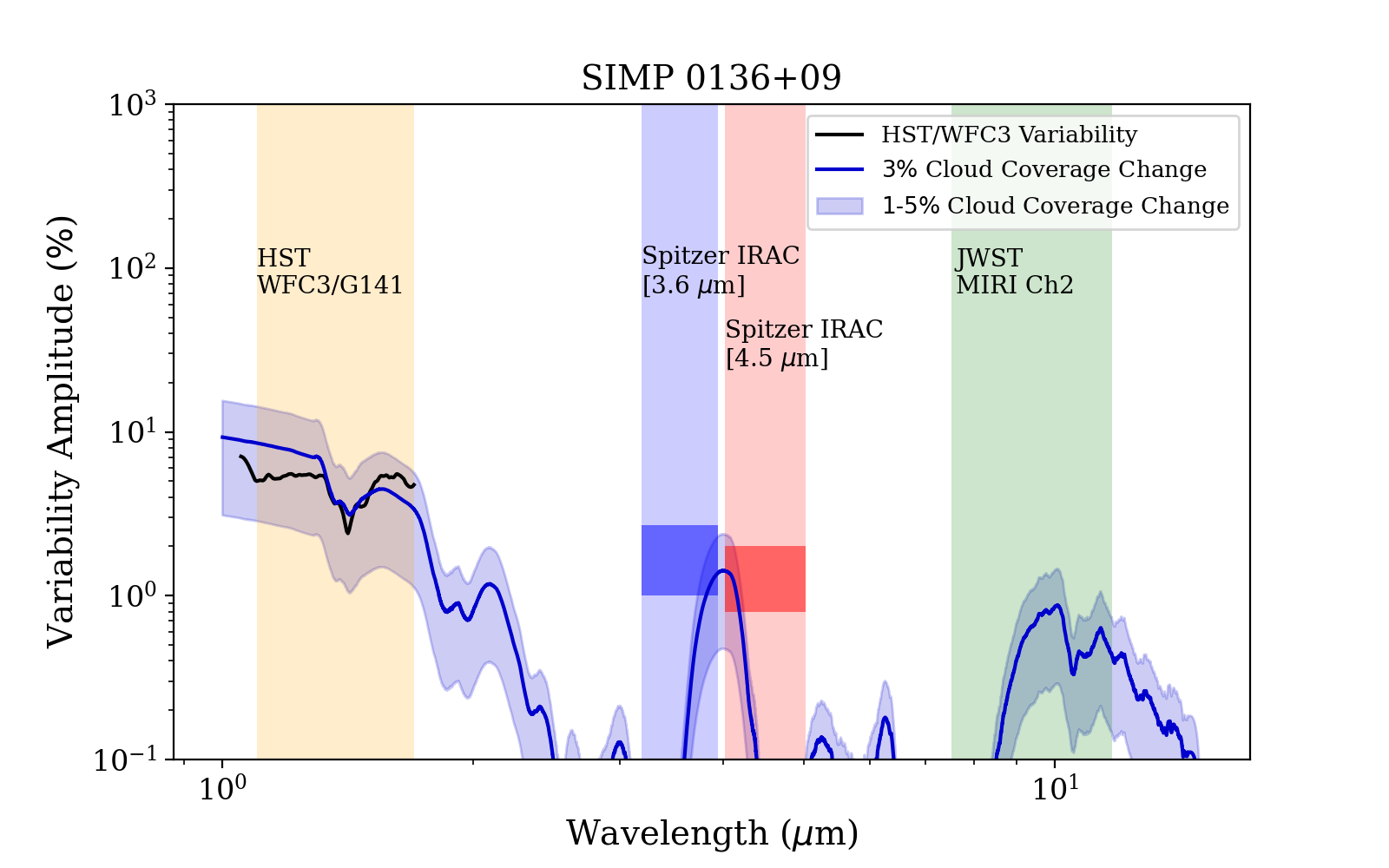}
   \includegraphics[width=0.75\textwidth]{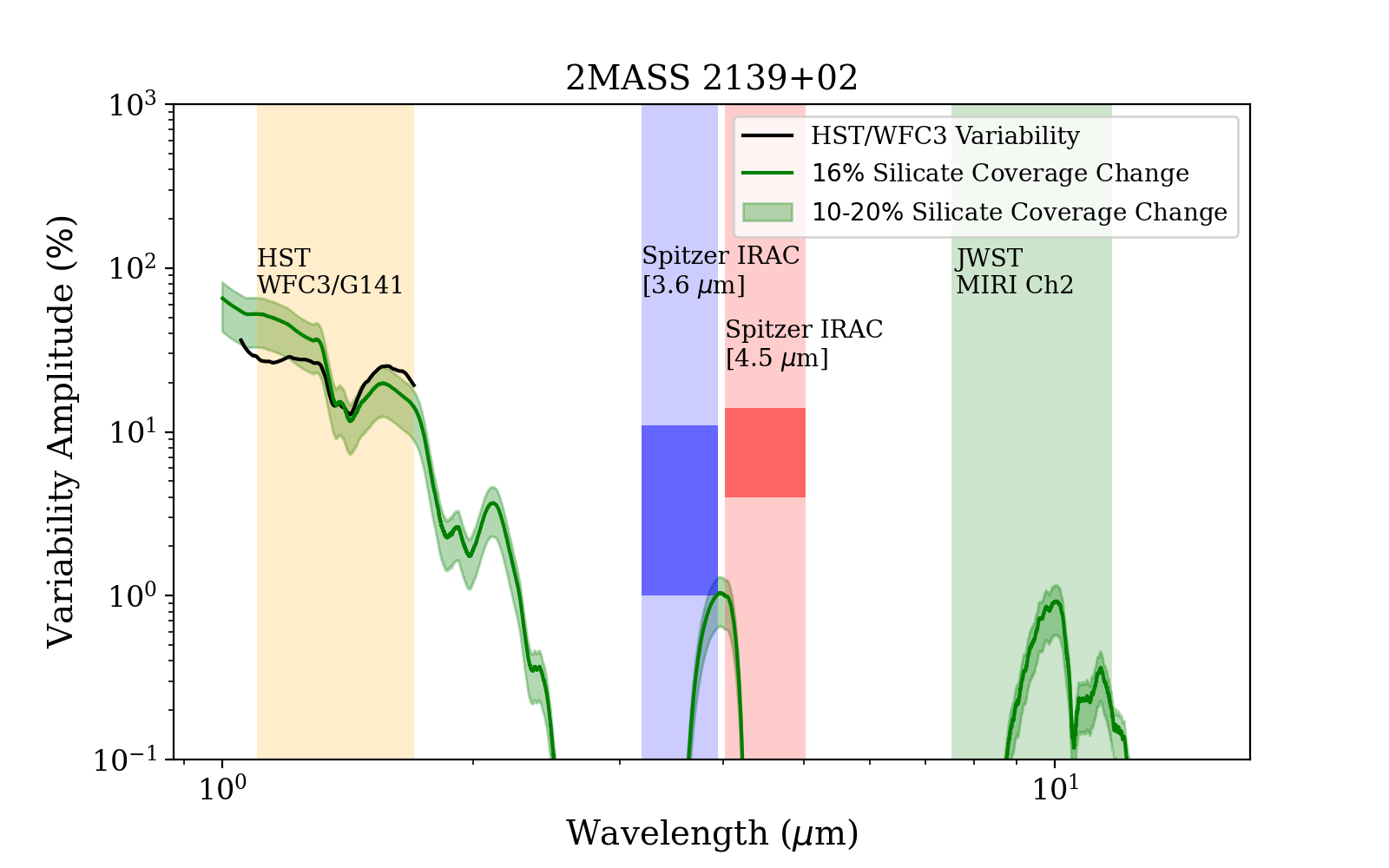}\\
   \caption{Comparison of our retrieval results with the observed spectral variability reported by \citet{Apai2013} for \obj{s0136} (top) and \obj{2m2139} (bottom). The predicted spectral variability driven by our retrieved patchy forsterite cloud slab is shown by the blue line for \obj{s0136} and green line for \obj{2m2139}. The range of amplitudes that would be predicted for a range of other silicate coverage changes is also shown by the semi-transparent blue and green shading for context.  The observed spectral variability from \citet{Apai2013} is shown in black. The darker blue and red regions within Spitzer/IRAC Channels 1 and 2 show the range of amplitudes reported for \obj{s0136} and \obj{2m2139} by \citet{Yang2016} during non-contemporaneous observations. The predicted variability signatures from our retrieved model broadly matches the observed  HST/WFC3 grism spectral variability and predicts lower amplitudes in the Spitzer $3.6~\mu$m and $4.5~\mu$m bands. Our retrieval also suggests that significant spectral variability will be observed in the silicate feature at $\sim10~\mu$m (accessible by JWST) due to the patchy forsterite slab.}
   \label{fig:variability_prediction}
\end{figure*}

To investigate this possibility directly, we estimate the spectral variability driven by a patchy forsterite cloud within our retrieved model by comparing the spectra from each patch. Additionally, we can use the observed HST/WFC3 variability reported by \citet{Apai2013} to estimate how the variability is related to the coverage fraction for each object -- a large change in silicate cloud coverage will naturally lead to higher amplitudes and vice versa. Treating our retrieved silicate cloud coverage percentages (shown in Table \ref{tab:cloudprops}) as the average silicate coverage for each target, we experimented with different silicate coverage changes until the retrieved variability signatures matched the observed values in the HST/WFC3 wavelength region of $1.1-1.7~\mu$m, and show our results in Figure \ref{fig:variability_prediction}. 
Using this technique, we find that for \obj{2m2139}, we can approximately reproduce the observed variability with a silicate cloud coverage percentage change of $16\%$. Assuming that the silicate cloud coverage constrained by our retrieval represents the average coverage percentage, our analysis predicts that the silicate cloud coverage changes from   $75-91\%$ to reproduce the observed variability. For \obj{s0136} we can reproduce the observed variability amplitude with a silicate cloud coverage percentage change of $3\%$.  Taking our retrieved coverage as the average value, we find that the silicate cloud coverage percentage likely changes from $69\%$ to $72\%$. Our results suggest that a greater change in silicate cloud coverage drives the larger variability amplitude observed in \obj{2m2139} compared to \obj{s0136}.

A number of studies have suggested a link between rotation rate and the size of cloud structures in extrasolar atmospheres. These results represent an excellent test case for this potential link. 
Atmospheric circulation models presented by \citet{Tan2021} show that the atmospheres of giant exoplanet and brown dwarfs are likely to be dominated by cloud-forming and clear-sky vortices that are seen to evolve with time.
The study highlights the importance of the Rossby deformation radius in setting the maximum size of storms and vortices. The Rossby deformation radius is inversely proportional to the local rotational rate when other parameters are the same. In the case of \obj{s0136} and \obj{2m2139}, our retrieval results have shown that their atmospheres are almost identical. Thus, any differences in the cloud structure may be driven by rotation rate. With a rotation period of $7.6~$hr, the deformation radius of \obj{2m2139} would be larger than that of \obj{s0136}, whose rotation period is $2.4~$hr. This would result in larger cloud-forming and clear-sky vortices emerging for \obj{2m2139}, which would naturally result in larger variability amplitudes as these large-scale structures rotate in and out of view. The results presented in this work strongly support this interpretation. Since the cloud layers for \obj{s0136} and \obj{2m2139} are at almost identical pressure levels, the changes in pressure and temperature achieved by seeing to pressures deeper than the forsterite slab are also very similar (i.e. see Figure \ref{fig:variability_pressure_temperature}). This means that greater contrasts between the forsterite and iron cloud layers do not drive greater amplitudes for \obj{2m2139}. Instead, it seems that larger atmospheric features are responsible for the large amplitudes in \obj{2m2139}.

We find that the variability predicted by our retrieval -- i.e. changes in brightness driven by the patchy forsterite slab cloud -- broadly matches the behaviour of the HST/WFC3 results. While the retrieval does not perfectly match the observed variability, it is expected that the variability behavior may have changed slightly between the HST/WFC3 observations and the epochs of our spectra. 
Moreover, the results shown in Figure \ref{fig:variability_prediction} also predict significant variability in the \textit{Spitzer} [$3.6~\mu$m] and [$4.5~\mu$m] bands, such as those reported by \citet{Yang2016}.  We show the non-contemporaneous \textit{Spitzer} variability amplitudes reported by \citet{Yang2016} in Figure \ref{fig:variability_prediction}, but do not use these values to scale our retrieval prediction. These variability amplitudes are consistent with the predicted amplitudes from our retrieval for \obj{s0136}, but are higher than the amplitude predicted for \obj{2m2139} in \textit{Spitzer} Channel 2 ([$4.5~\mu$m]) The light curves of \obj{s0136} and \obj{2m2139} reported by \citet{Yang2016} show significant light curve evolution, and hence a large range in amplitudes that is reflected in Figure \ref{fig:variability_prediction}. This shows the importance of simultaneous variability monitoring in order to investigate extrasolar atmospheres in detail.

Finally, our retrieval model predicts significant variability at $10~\mu$m that is driven by the silicate scattering feature. Although multiple studies have speculated that variability is likely present at this feature \citep[e.g.][]{Luna2021}, variability monitoring at these wavelengths has not been possible until now. JWST/MIRI will provide the means to monitor this region closely in order to identify silicate clouds as the driver of variability. 

This broad agreement is promising, but there remain some details that that retrieval results do not explain. In particular, using simultaneous variability monitoring with HST and Spitzer, \citet{Yang2016} report a phase shift of $30^{\circ}$ between the near-IR and mid-IR light curves and a correlation between these measured phase shifts and the pressure levels probed by each wavelength. 
 Such behavior is likely driven by complex vertical behaviour, such as vertically varying cloud structures or P/T profile variations above the clouds that are not captured by our 1-dimensional patchy cloud framework.

\subsection{C/O Ratio Among Two Carina-Near Members}
The chemical composition of extrasolar atmospheres is often considered the key to disentangling the formation pathways of stars, brown dwarfs and planets. In particular, a large number of studies have targeted the C/O ratio in an attempt to reveal the formation mechanism of exoplanets \citep{Oberg2011, Madhusudhan2012}, however it is likely to be difficult to tease out a direct link between measured C/O and formation mechanism \citep{Molliere2022}. Measuring C/O ratios in benchmark systems is thus critical for understanding and calibrating this measurement.

\obj{s0136} and \obj{2m2139} are both members of the 200~Myr Carina-Near moving group and therefore are expected to have formed in the same molecular cloud, alongside other stars, brown dwarfs and planetary-mass objects. With similar masses, we would also expect similar subsequent evolution. To first order, we would thus expect their C/O ratios to be consistent with each other, and with the larger sample of Carina-Near brown dwarfs and stars. 

Numerous studies have highlighted the difference between the \textit{atmospheric} C/O ratio and the \textit{intrinsic} C/O ratio \citep{Line2015, Molliere2022,Calamari2022}. \textit{Atmospheric} C/O ratios can be measured to high precision using atmospheric retrievals \citep[e.g.][]{Zalesky2019, Calamari2022}, however how they relate to the \textit{intrinsic} values is an open question. Cloud condensation processes, such as the formation of a forsterite cloud found in this work, can deplete atmospheric oxygen or other species, resulting in atmospheric C/O ratios that differ from their intrinsic values.
Within our standard retrieval framework we estimate the atmospheric C/O ratio for our targets by assuming that all carbon and oxygen are contained in the considered absorbing gases. 
We find consistent C/O ratios in our targets (see corner plots in Figures \ref{fig:0136_gascorner} and \ref{fig:2139_gascorner}) -- measuring C/O$=0.79\pm0.02$ for \obj{s0136} and C/O$=0.82\pm0.03$ for \obj{2m2139}. Their agreement is in line with our first-order expectations that these two worlds formed from the same starting material,  proceeded on a similar evolutionary pathway and have very similar atmospheres.

While their atmospheric C/O ratios are in agreement, it is challenging to estimate the intrinsic or bulk C/O ratios. Our C/O estimate does not include any oxygen contained in condensates such as Mg$_2$SiO$_4$, which we have detected in the atmospheres of our targets. Atmospheric retrieval studies in the literature have typically applied corrections to account for such oxygen depletion. \citet{Burrows1999} estimate that  3.28 oxygen atoms are removed from the gas phase for every silicon atom under the assumption that enstatite, (MgSiO$_3$) is the dominant condensation species with  forsterite (Mg$_2$SiO$_4$) as a secondary species. Since we find that forsterite is the dominant condensate in the atmospheres of \obj{s0136} and \obj{2m2139}, this would result in the depletion of up to 4 oxygen atoms per silicon atom. Taking this into account, we estimate that the \textit{intrinsic} C/O ratio could be as low as $0.56$ for \obj{s0136} and $0.59$ for \obj{2m2139}, values more in line with stellar C/O ratios reported in the literature \citep{ Nissen2013, Nissen2015,Brewer2016}.

If we assume that both \obj{s0136} and \obj{2m2139} formed like stars, we would expect that their intrinsic C/O ratios match those of their stellar counterparts in the Carina-Near moving group. Measurements of stellar and brown dwarf C/O ratios across well-characterized young moving groups such as Carina-Near would thus be an opportunity to gain a greater understanding of the relationship between atmospheric and intrinsic C/O ratios in cloudy worlds.


\begin{figure*}[tb]
   \centering
   \includegraphics[width=0.47\textwidth]{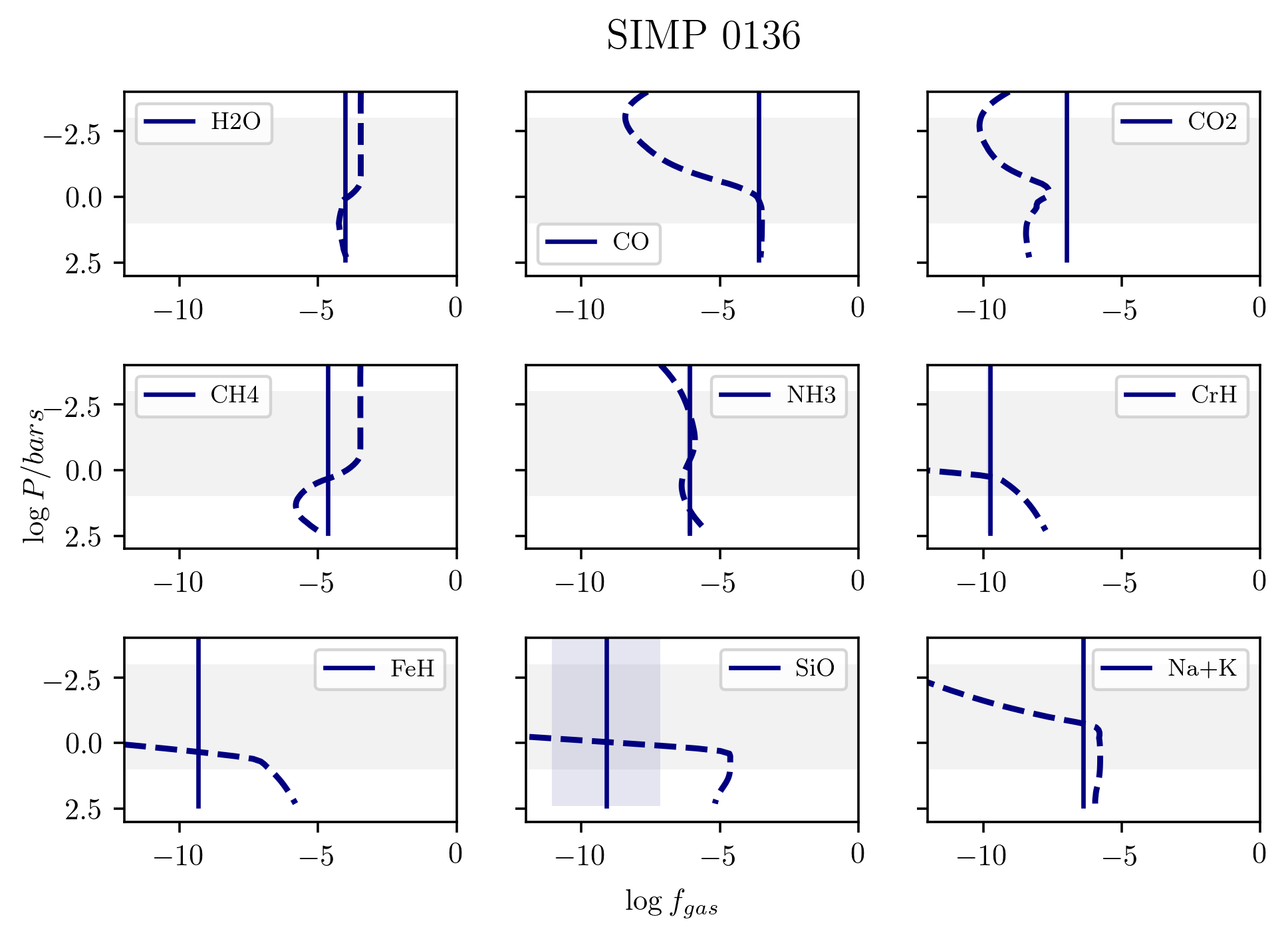}
     \includegraphics[width=0.47\textwidth]{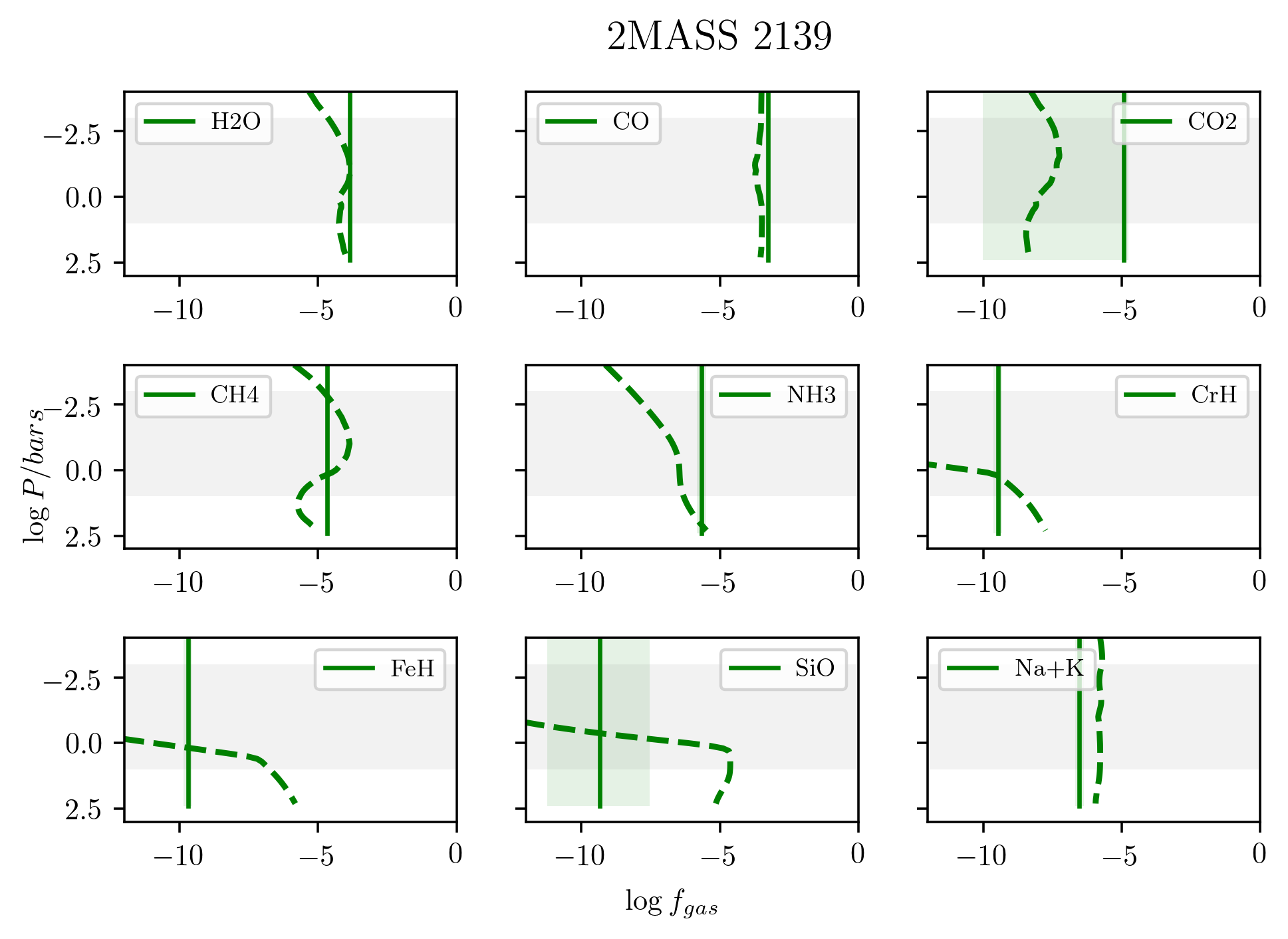}
   \caption{Retrieved gas fractions compared to predictions from thermochemical grids for \obj{s0136} (left) and \obj{2m2139} (right). Equilibrium predictions are shown as dashed lines and are calculated for our estimated [M/H] and C/O values. The solid straight lines and shading show hour median retrieved values and 16th to 84th percentiles, respectively.}
   \label{fig:abundances_comparison}
\end{figure*}

\subsection{Differences in Chemistry}
We present our retrieved gas abundances in Table \ref{tab:gases}. The retrieved gas abundances show perhaps the greatest differences between \obj{s0136} and \obj{2m2139}. 
Of particular interest are the lower abundances retrieved for H$_2$O and CO for \obj{s0136} as these are some of the most abundant gases found for \obj{2m2139} and are significantly different in our two targets. These lower abundances are responsible for the low metallicity that we derive for \obj{s0136}. 
These differences may be due to our data quality and/or retrieval methodology, or may have an astrophysical explanation. We will briefly touch on these possibilities in this section.

As discussed in Section \ref{sec:spectra}, the available data for \obj{s0136} and \obj{2m2139} are quite different. In particular, for \obj{s0136}, the combination of SpeX, AKARI and Spitzer/IRS data provides us with complete wavelength coverage from $1-15~\mu$m, while for \obj{2m2139} we are missing the $2.5-5~\mu$m region. This means that we achieve more complete coverage of the pressure levels within the atmosphere for \obj{s0136}. To test whether these discrepancies were caused by the inclusion of the AKARI data  for \obj{s0136}, we ran our highest ranked model on only the SpeX and Spitzer/IRS portions of the spectrum, and found that the retrieved gas abundances (along with the other retrieved parameters) were consistent with and without this portion of the spectrum. From this we conclude that the inclusion of the $2.5-5~\mu$m spectrum does not affect measured abundances in this case, and does not cause the low metallicity derived for \obj{s0136}.

As discussed in Section \ref{sec:spectra}, the SpeX spectra for both targets are of similar quality, with average SNR ranging from 167-180. The SNR of the Spitzer/IRS spectrum of \obj{2m2139} is significantly lower than that of \obj{s0136} (average SNR of 5 versus 15). SNR differences are unlikely to cause the lower measured abundances of H$_2$O and CO in \obj{s0136} since our data for this target is of higher SNR, thus making it easier to detect such gases in the spectra.

Another possibility is that our retrieval framework may be responsible for the differing abundance measurements. 
The \textit{Brewster} retrieval framework uses vertically constant mixing ratios to describe each gas within the model. 
In Figure \ref{fig:abundances_comparison} we show our retrieved gas mixing ratios along with predictions from thermochemical equilibrium models interpolated for our derived metallicity and C/O ratio. Thermochemical equilibrium models predict that many of the included gases will vary substantially as a function of altitude within the pressure levels probed by our spectra (e.g. FeH, CrH), while others are predicted to remain relatively constant with altitude. Our gases of interest, H$_2$O and CO, are not predicted to change significantly, but it is possible that by failing to capture the behavior of gases that do change substantially, this could have an indirect effect on the retrieved abundances of H$_2$O and CO. However, we expect that this would affect the retrieved abundances for both targets in the same way, so our assumption of vertically constant mixing ratios cannot be the underlying cause for these changes. Rather, it is likely that there are differences in either the data (as discussed previously) or astrophysical differences (discussed next) that are revealed by these differing abundance measurements.

There are a number of astrophysical processes that may drive these differing abundances. The first major difference between our targets is that \obj{s0136} has been shown to be an auroral emitter \citep{Kao2016} while there is no evidence in the literature that \obj{2m2139} emits aurorae. 

\begin{figure*}
   \centering
       \includegraphics[width=0.85\textwidth]{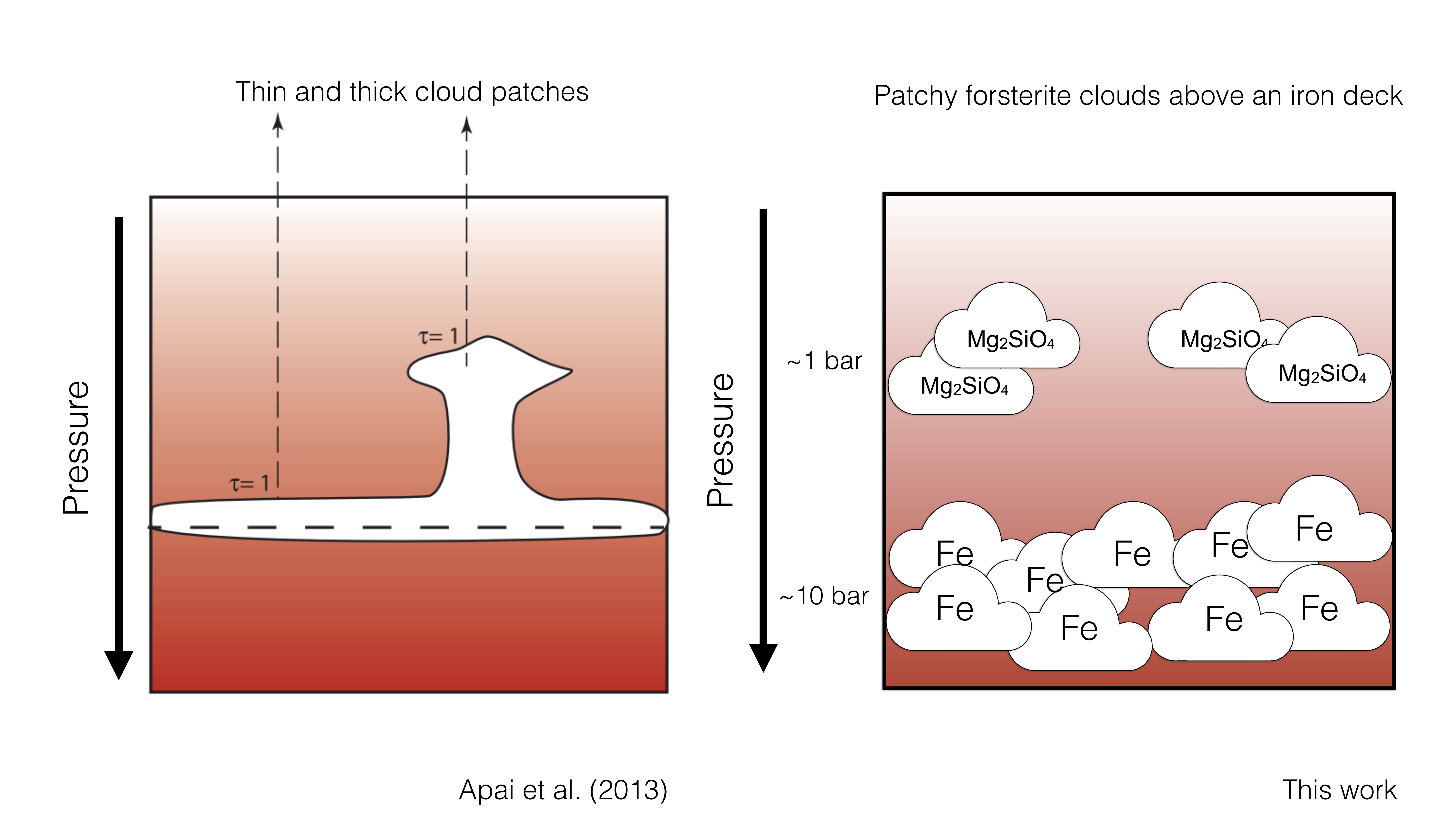}
   \caption{Illustrations proposed by \citet{Apai2013} (left) and this work (right) for the atmospheric structure of \obj{s0136} and \obj{2m2139}. \citet{Apai2013} find that a cloud deck of varying thickness could explain their observed HST/WFC3 variability observations. Our atmospheric retrieval analysis of a single $1-15~\mu$m spectrum provides a similar, but more detailed view into the atmospheres of \obj{s0136} and \obj{2m2139}.}
   \label{fig:illustration}
\end{figure*}

Auroral activity in brown dwarfs is of the same nature as the auroral emission produced by the giant planets in our own solar system and is driven by strong-field-aligned currents that drive accelerated electron beams. These accelerated electrons can lead to the onset of the electron cyclotron maser instability (ECMI) that produces detectable radio emission \citep[e.g.][]{Hallinan2015, Kao2016,Kao2018, Pineda2017,Allers2020,Richey-Yowell2020}. These energetic electron beams may also have a significant effect on the atmosphere. In Jupiter and Saturn, collisions between electrons and the atmospheric gases leads to optical and UV emissions due to excitation and ionization processes and subsequent ion chemistry leads to the formation of  strongly emitting H$_3^+$ \citep{Perry1999,Vasavada1999} ions. It is likely that auroral activity in the atmospheres of brown dwarfs also leads to the formation of detectable amounts of H$_3^+$, but this has not yet been observed \citep{Pineda2017, Gibbs2022}. In this work we can speculate on whether auroral activity on \obj{s0136} could lead to the suppression of H$_2$O and CO absorption. 

One possibility is that the presence of auroral currents leads to the destruction of H$_2$O and CO. Interestingly, \citet{Helling2019} provide such a mechanism. Using three-dimensional simulations that include kinetic cloud formation and kinetic ion-neutral chemistry, the authors model the atmospheric chemistry of the brown dwarf auroral emitter LSR$-$1835 to study the formation and evolution of H$_3^+$. These simulations show that for the case of LSR$-$1835, small amounts of H$_3^+$ ions can form at pressures down to $\sim0.1~$bar, but that H$_3^+$  reacts rapidly with H$_2$O and CO, and can produce the relatively stable ion hydronium, H$_3$O$^+$, at pressures down to $\sim1~$bar. \citet{Gibbs2022} recently provided upper limits on the abundances of H$_3^+$ in the atmosphere of \obj{s0136}, suggesting that this ion may indeed be short-lived in the atmosphere. Whether the reactions of H$_2$O and CO with H$_3^+$ could account for their significantly depleted abundances in \obj{s0136} will be an interesting future investigation.

A second major difference is the differing rotation periods of our targets -- $2.4~$hr for \obj{s0136} and $7.6~$hr for \obj{2m2139}. Atmospheric dynamical simulations presented by \citet{Tan2021} highlight the importance of rotation rate on regulating the atmospheres of brown dwarfs and giant planets. One of their conclusions is that faster rotation tends to weaken the vertical transport of vapor and clouds, which may have a measurable effect on retrieved abundances of various species. However, it is hard to explain the relative depletion of H$_2$O and CO in \obj{s0136} with this mechanism while other important and abundant gases such as CH$_4$ remain consistent between \obj{s0136} and \obj{2m2139}. In-depth atmospheric dynamical simulations of \obj{s0136} and \obj{2m2139} are needed to fully explore the influence of rotation period on the abundances of atmospheric gases.

Finally, there is likely to be some intrinsic atmospheric diversity among extrasolar atmospheres. Carrying out detailed retrieval studies on a larger sample of objects will reveal the range of atmospheric parameters that we can expect from otherwise similar extrasolar worlds.

\subsection{Atmospheric Retrievals in the Era of JWST}

Retrievals of extrasolar atmospheres utilizing high precision spectra with wide wavelength ranges of $1-20~\mu$m will soon be commonplace in the era of JWST \citep[e.g.][]{Miles2022}. The results of this work, alongside the previously published retrieval study of \obj{2m2224} by \citet{Burningham2021}, which both make use of Spitzer/IRS data to achieve a wavelength range of $1-15~\mu$m,  highlight the detail in which we can characterize an atmosphere with such data. Both studies found that the data were best described by a complex cloud structure -- slabs containing enstatite and quartz along with an iron deck for \obj{2m2224} and a patchy forsterite slab with an iron deck for both \obj{s0136} and \obj{2m2139}. Since JWST will provide spectra of higher resolution and higher signal-to-noise, that are acquired near-simultaneously, retrievals of these data will likely reveal the atmospheres of these worlds in even more detail and accuracy than we have presented here.


Additionally, JWST will allow us to extend our spectroscopic monitoring efforts beyond the HST/WFC3 grism wavelength range of $1.1-1.7~\mu$m, and to provide a wealth of information on how wavelengths beyond $2~\mu$m change over time. Spectroscopic variability studies at a wider wavelength range will reveal the time-resolved behavior of these atmospheres at a broader range of pressures than before. Of particular interest will be the behavior of the $10~\mu$m silicate feature \citep[e.g.][Figure \ref{fig:variability_prediction} of this work]{Luna2021}. Spectroscopic variability monitoring with JWST may even enable longitudinally-resolved retrievals for brown dwarfs and planetary-mass objects, as has recently been carried out for hot Jupiters \citep[e.g.][]{Cubillos2021,MacDonald2022,Nixon2022}. The work presented here represents a good starting point for future retrieval studies using wide wavelength coverage provided by JWST.

\section{Summary and Conclusions}

In this work we present an atmospheric retrieval analysis of $1-15~\mu$m spectra of the isolated exoplanet analogs \obj{s0136} and \obj{2m2139} using the \textit{Brewster} retrieval framework. These targets are an interesting pair of objects to study, since they share their ages, masses, temperatures and variability properties with each other.

We test a number of different models and find that the spectra of \obj{s0136} and \obj{2m2139} are best described by a model containing a patchy forsterite slab cloud lying above an iron deck cloud. The properties of the clouds are strikingly similar between our two targets. Both atmospheres contain a patchy forsterite slab whose cloud base is located at a pressure of $1.3-1.7~$bar that is composed of sub-micron grains and an  iron deck that is located at a pressure of $\sim7~$bar that is also composed of sub-micron sized grains. We constrain the covering fractions for the forsterite clouds to be $\sim0.7$ for \obj{s0136} and $\sim0.8$ for \obj{2m2139}. 

Both \obj{s0136} and \obj{2m2139} are known to exhibit large amplitude spectroscopic variability \citep{Artigau2009,Radigan2012,Apai2013} which has generally been interpreted as arising from inhomogeneous clouds rotating in and out of view. 
The preference of our retrieval analysis for models that contain patchy clouds provides independent support for inhomogeneous clouds in the atmospheres of our targets. We investigate whether our retrieved model atmosphere is consistent with the spectroscopic variability observations presented by \citet{Apai2013} and find that the spectral behavior of the variability broadly matches the predicted variability that we can derive from our retrieval. We find that the amplitude of variability can be reproduced if the silicate cloud coverage changes from $69-72\%$ for \obj{s0136} and $75-91\%$ for \obj{2m2139}.  This work  suggests that the larger variability amplitudes observed for \obj{2m2139} are driven by larger atmospheric features that form as a result of its larger Rossby deformation radius.  Our retrieved atmospheric model also predicts significant variability in the [$3.6~\mu$m] and [$4.5~\mu$m] Spitzer bands \citep[as has been observed by ][]{Yang2016} and across the $10~\mu$m scattering feature, which can be probed with JWST/MIRI.

In Figure \ref{fig:illustration}, we provide a comparison that summarizes the atmospheric structure presented by \citet{Apai2013} (left) and in this work (right). Based on HST/WFC3 spectroscopic variability observations, \citet{Apai2013} proposed that the atmosphere consists of a cloud of varying thickness, with regions composed of hotter, thinner clouds and regions of cooler, thicker clouds. Using a spectrum for each object from $1-15~\mu$m , the work presented in this study clarifies this view significantly. Our retrieval analysis finds evidence for two distinct cloud layers, the top cloud composed of patchy forsterite slab clouds and the bottom cloud composed of a thick iron deck. We have also  constrained the pressure of each cloud layer as well as the particles sizes for each cloud species.

We find consistent atmospheric C/O ratios of $0.79\pm0.02~$dex for \obj{s0136} and $0.82\pm0.03~$dex for \obj{2m2139}, which is expected given that they are both members of the Carina-Near moving group with very similar atmospheres. Estimating their intrinsic C/O ratios is challenging because of oxygen depletion due to forsterite clouds forming in the atmosphere, but we estimate intrinsic C/O ratios as low as $0.56$ for \obj{s0136} and $0.59$ for \obj{2m2139}, that are closer to expectations for stars of similar ages.

We find slightly different abundances of H$_2$O and CO in our targets, which may be due to the data quality, retrieval framework and/or astrophysical processes such as auroral activity or atmospheric dynamics driven by their different rotation periods. As we build upon the sample of brown dwarfs with in-depth atmospheric retrievals, these differences can be placed in context and the aforementioned possibilities can be investigated in more detail.

Finally, we believe that the results presented here represent an encouraging preliminary study for the types of investigations that will open up as observations of brown dwarfs and giant exoplanets at wavelengths of $1-20~\mu$m become commonplace \citep[e.g. see][for the first such JWST observation presented in the literature]{Miles2022}. The results shown in this work highlight the power of such wavelength ranges for characterizing extrasolar atmospheres in exceptional detail. As these current and upcoming JWST observations will be of higher signal-to-noise, higher resolution, and will be taken near-simultaneously, future retrieval studies that make use of these data should provide an even clearer view of extrasolar atmospheres than the one presented in this work.

\section{Acknowledgements}
{We thank the anonymous referee for useful comments which improved the quality of this work.} The data used in this publication were collected through the MENDEL high performance computing (HPC) cluster at the American Museum of Natural History. This HPC cluster was developed with National Science Foundation (NSF) Campus Cyberinfrastructure support through Award \#1925590. This work has made use of the University of Hertfordshire's high-performance computing facility. This work was supported by NSF Award 1909776, NASA XRP Award \#80NSSC22K0142  and  HST-GO-15924.001-A.  E.G. acknowledges support from the Heising-Simons Foundation for this research. C.V.M acknowledges support from NSF AAG grant number 1910969. This work benefited from the 2022 Exoplanet Summer Program in the Other Worlds Laboratory (OWL) at the University of California, Santa Cruz, a program funded by the Heising-Simons Foundation.


\bibliography{main}{}

\begin{thebibliography}{}
\expandafter\ifx\csname natexlab\endcsname\relax\def\natexlab#1{#1}\fi
\providecommand{\url}[1]{\href{#1}{#1}}
\providecommand{\dodoi}[1]{doi:~\href{http://doi.org/#1}{\nolinkurl{#1}}}
\providecommand{\doeprint}[1]{\href{http://ascl.net/#1}{\nolinkurl{http://ascl.net/#1}}}
\providecommand{\doarXiv}[1]{\href{https://arxiv.org/abs/#1}{\nolinkurl{https://arxiv.org/abs/#1}}}

\bibitem[{Ackerman \& Marley(2001)}]{Ackerman2001}
Ackerman, A.~S., \& Marley, M.~S. 2001, Astrophysical Journal, 765, 872

\bibitem[{Allard {et~al.}(2012)Allard, Homeier, \& Freytag}]{Allard2012}
Allard, F., Homeier, D., \& Freytag, B. 2012, Philosophical Transactions of the
  Royal Society A: Mathematical, Physical and Engineering Sciences, 370, 2765

\bibitem[{{Allers} {et~al.}(2020){Allers}, {Vos}, {Biller}, \&
  {Williams}}]{Allers2020}
{Allers}, K.~N., {Vos}, J.~M., {Biller}, B.~A., \& {Williams}, P. K.~G. 2020,
  Science, 368, 169

\bibitem[{Apai {et~al.}(2013)Apai, Radigan, Buenzli, Burrows, Reid, \&
  Jayawardhana}]{Apai2013}
Apai, D., Radigan, J., Buenzli, E., {et~al.} 2013, The Astrophysical Journal,
  768, 121

\bibitem[{Apai {et~al.}(2017)Apai, Karalidi, Marley, Yang, Flateau, Metchev,
  Cowan, Buenzli, Burgasser, Radigan, Artigau, \& Lowrance}]{Apai2017}
Apai, D., Karalidi, T., Marley, M.~S., {et~al.} 2017, Science, 357, 683

\bibitem[{Artigau {et~al.}(2009)Artigau, Bouchard, Doyon, \&
  Lafreni{\`{e}}re}]{Artigau2009}
Artigau, {\'{E}}., Bouchard, S., Doyon, R., \& Lafreni{\`{e}}re, D. 2009, The
  Astrophysical Journal, 701, 1534

\bibitem[{{Artigau} {et~al.}(2006){Artigau}, {Doyon}, {Lafreni{\`e}re},
  {Nadeau}, {Robert}, \& {Albert}}]{Artigau2006}
{Artigau}, {\'E}., {Doyon}, R., {Lafreni{\`e}re}, D., {et~al.} 2006, \apjl,
  651, L57

\bibitem[{Asplund {et~al.}(2009)Asplund, Grevesse, Sauval, \&
  Scott}]{Asplund2009}
Asplund, M., Grevesse, N., Sauval, A.~J., \& Scott, P. 2009, Annual Review of
  Astronomy and Astrophysics, 47, 481

\bibitem[{Barman {et~al.}(2011)Barman, Macintosh, Konopacky, \&
  Marois}]{Barman2011}
Barman, T.~S., Macintosh, B., Konopacky, Q.~M., \& Marois, C. 2011, The
  Astrophysical Journal, 735, L39

\bibitem[{{Barstow} \& {Heng}(2020)}]{Barstow2020}
{Barstow}, J.~K., \& {Heng}, K. 2020, \ssr, 216, 82

\bibitem[{{Bell}(1980)}]{Bell1980}
{Bell}, K.~L. 1980, Journal of Physics B Atomic Molecular Physics, 13, 1859

\bibitem[{{Bell} \& {Berrington}(1987)}]{Bell1987}
{Bell}, K.~L., \& {Berrington}, K.~A. 1987, Journal of Physics B Atomic
  Molecular Physics, 20, 801

\bibitem[{Biller {et~al.}(2015)Biller, Vos, Bonavita, Buenzli, Baxter,
  Crossfield, Allers, Liu, Bonnefoy, Deacon, Brandner, Schlieder, Dupuy,
  Kopytova, Manjavacas, Allard, Homeier, \& Henning}]{Biller2015}
Biller, B.~A., Vos, J., Bonavita, M., {et~al.} 2015, Astrophysical Journal
  Letters, 813, 1

\bibitem[{Biller {et~al.}(2018)Biller, Vos, Buenzli, Allers, Bonnefoy, Charnay,
  B{\'{e}}zard, Allard, Homeier, Bonavita, Brandner, Crossfield, Dupuy,
  Henning, Kopytova, Liu, Manjavacas, \& Schlieder}]{Biller2018}
Biller, B.~A., Vos, J., Buenzli, E., {et~al.} 2018, The Astronomical Journal,
  155, 95

\bibitem[{{Bohn} {et~al.}(2020){Bohn}, {Kenworthy}, {Ginski}, {Rieder},
  {Mamajek}, {Meshkat}, {Pecaut}, {Reggiani}, {de Boer}, {Keller}, {Snik}, \&
  {Southworth}}]{Bohn2020}
{Bohn}, A.~J., {Kenworthy}, M.~A., {Ginski}, C., {et~al.} 2020, \apjl, 898, L16

\bibitem[{{Bohn} {et~al.}(2021){Bohn}, {Ginski}, {Kenworthy}, {Mamajek},
  {Pecaut}, {Mugrauer}, {Vogt}, {Adam}, {Meshkat}, {Reggiani}, \&
  {Snik}}]{Bohn2021}
{Bohn}, A.~J., {Ginski}, C., {Kenworthy}, M.~A., {et~al.} 2021, \aap, 648, A73

\bibitem[{{Bowler} {et~al.}(2020){Bowler}, {Zhou}, {Morley}, {Kataria},
  {Bryan}, {Benneke}, \& {Batygin}}]{Bowler2020}
{Bowler}, B.~P., {Zhou}, Y., {Morley}, C.~V., {et~al.} 2020, \apjl, 893, L30

\bibitem[{{Brewer} \& {Fischer}(2016)}]{Brewer2016}
{Brewer}, J.~M., \& {Fischer}, D.~A. 2016, \apj, 831, 20

\bibitem[{{Brock} {et~al.}(2021){Brock}, {Barman}, {Konopacky}, \&
  {Stone}}]{Brock2021}
{Brock}, L., {Barman}, T., {Konopacky}, Q.~M., \& {Stone}, J.~M. 2021, \apj,
  914, 124

\bibitem[{Buenzli {et~al.}(2014)Buenzli, Apai, Radigan, Reid, \&
  Flateau}]{Buenzli2014}
Buenzli, E., Apai, D., Radigan, J., Reid, I.~N., \& Flateau, D. 2014, The
  Astrophysical Journal, 782, 77

\bibitem[{Buenzli {et~al.}(2015)Buenzli, Marley, Apai, Saumon, Biller,
  Crossfield, \& Radigan}]{Buenzli2015}
Buenzli, E., Marley, M.~S., Apai, D., {et~al.} 2015, The Astrophysical Journal,
  812, 163

\bibitem[{{Burgasser} {et~al.}(2006){Burgasser}, {Geballe}, {Leggett},
  {Kirkpatrick}, \& {Golimowski}}]{Burgasser2006}
{Burgasser}, A.~J., {Geballe}, T.~R., {Leggett}, S.~K., {Kirkpatrick}, J.~D.,
  \& {Golimowski}, D.~A. 2006, \apj, 637, 1067

\bibitem[{{Burgasser} {et~al.}(2008){Burgasser}, {Liu}, {Ireland}, {Cruz}, \&
  {Dupuy}}]{Burgasser2008b}
{Burgasser}, A.~J., {Liu}, M.~C., {Ireland}, M.~J., {Cruz}, K.~L., \& {Dupuy},
  T.~J. 2008, \apj, 681, 579

\bibitem[{{Burningham} {et~al.}(2017){Burningham}, {Marley}, {Line}, {Lupu},
  {Visscher}, {Morley}, {Saumon}, \& {Freedman}}]{Burningham2017}
{Burningham}, B., {Marley}, M.~S., {Line}, M.~R., {et~al.} 2017, \mnras, 470,
  1177

\bibitem[{{Burningham} {et~al.}(2021){Burningham}, {Faherty}, {Gonzales},
  {Marley}, {Visscher}, {Lupu}, {Gaarn}, {Bieger}, {Freedman}, \&
  {Saumon}}]{Burningham2021}
{Burningham}, B., {Faherty}, J.~K., {Gonzales}, E.~C., {et~al.} 2021, \mnras

\bibitem[{Burrows {et~al.}(2001)Burrows, Hubbard, Lunine, \&
  Liebert}]{Burrows2001}
Burrows, A., Hubbard, W.~B., Lunine, J.~I., \& Liebert, J. 2001, Reviews of
  Modern Physics, 73, 719

\bibitem[{{Burrows} \& {Sharp}(1999)}]{Burrows1999}
{Burrows}, A., \& {Sharp}, C.~M. 1999, \apj, 512, 843

\bibitem[{{Burrows} {et~al.}(2006){Burrows}, {Sudarsky}, \&
  {Hubeny}}]{Burrows2006}
{Burrows}, A., {Sudarsky}, D., \& {Hubeny}, I. 2006, \apj, 640, 1063

\bibitem[{{Burrows} \& {Volobuyev}(2003)}]{Burrows2003}
{Burrows}, A., \& {Volobuyev}, M. 2003, \apj, 583, 985

\bibitem[{Burrows {et~al.}(1997)Burrows, Marley, Hubbard, Lunine, Guillot,
  Saumon, Freedman, Sudarsky, \& Sharp}]{Burrows1997}
Burrows, A., Marley, M., Hubbard, W.~B., {et~al.} 1997, The Astrophysical
  Journal, 491, 856

\bibitem[{{Calamari} {et~al.}(2022){Calamari}, {Faherty}, {Burningham},
  {Gonzales}, {Bardalez-Gagliuffi}, {Vos}, {Gemma}, {Whiteford}, \&
  {Gaarn}}]{Calamari2022}
{Calamari}, E., {Faherty}, J.~K., {Burningham}, B., {et~al.} 2022, \apj, 940,
  164

\bibitem[{{Carter} {et~al.}(2022){Carter}, {Hinkley}, {Kammerer}, {Skemer},
  {Biller}, {Leisenring}, {Millar-Blanchaer}, {Petrus}, {Stone}, {Ward-Duong},
  {Wang}, {Girard}, {Hines}, {Perrin}, {Pueyo}, {Balmer}, {Bonavita},
  {Bonnefoy}, {Chauvin}, {Choquet}, {Christiaens}, {Danielski}, {Kennedy},
  {Matthews}, {Miles}, {Patapis}, {Ray}, {Rickman}, {Sallum}, {Stapelfeldt},
  {Whiteford}, {Zhou}, {Absil}, {Boccaletti}, {Booth}, {Bowler}, {Chen},
  {Currie}, {Fortney}, {Grady}, {Greenbaum}, {Henning}, {Hoch}, {Janson},
  {Kalas}, {Kenworthy}, {Kervella}, {Kraus}, {Lagage}, {Liu}, {Macintosh},
  {Marino}, {Marley}, {Marois}, {Matthews}, {Mawet}, {McElwain}, {Metchev},
  {Meyer}, {Molliere}, {Moran}, {Morley}, {Mukherjee}, {Pantin}, {Quirrenbach},
  {Rebollido}, {Ren}, {Schneider}, {Vasist}, {Worthen}, {Wyatt},
  {Briesemeister}, {Bryan}, {Calissendorff}, {Cantalloube}, {Cugno}, {De
  Furio}, {Dupuy}, {Factor}, {Faherty}, {Fitzgerald}, {Franson}, {Gonzales},
  {Hood}, {Howe}, {Kuzuhara}, {Lagrange}, {Lawson}, {Lazzoni}, {Lew}, {Liu},
  {Lloyd}, {Martinez}, {Mazoyer}, {Quanz}, {Adams Redai}, {Samland},
  {Schlieder}, {Tamura}, {Tan}, {Uyama}, {Vigan}, {Vos}, {Wagner}, {Wolff},
  {Ygouf}, {Zhang}, {Zhang}, \& {Zhang}}]{Carter2022}
{Carter}, A.~L., {Hinkley}, S., {Kammerer}, J., {et~al.} 2022, arXiv e-prints,
  arXiv:2208.14990

\bibitem[{Chauvin {et~al.}(2004)Chauvin, Lagrange, Dumas, Zuckerman, Mouillet,
  Song, Beuzit, \& Lowrance}]{Chauvin2004}
Chauvin, G., Lagrange, A., Dumas, C., {et~al.} 2004, Astronomy and
  Astrophysics, 425, L29

\bibitem[{{Cubillos} {et~al.}(2021){Cubillos}, {Keating}, {Cowan}, {Vos},
  {Burningham}, {Ygouf}, {Karalidi}, {Zhou}, \& {Gonzales}}]{Cubillos2021}
{Cubillos}, P.~E., {Keating}, D., {Cowan}, N.~B., {et~al.} 2021, \apj, 915, 45

\bibitem[{Cushing {et~al.}(2006)Cushing, Roellig, Marley, Saumon, Leggett,
  Kirkpatrick, Wilson, Sloan, Mainzer, {Van Cleve}, \& Houck}]{Cushing2006}
Cushing, M.~C., Roellig, T.~L., Marley, M.~S., {et~al.} 2006, The Astrophysical
  Journal, 648, 614

\bibitem[{{Cutri} {et~al.}(2003){Cutri}, {Skrutskie}, {van Dyk}, {Beichman},
  {Carpenter}, {Chester}, {Cambresy}, {Evans}, {Fowler}, {Gizis}, {Howard},
  {Huchra}, {Jarrett}, {Kopan}, {Kirkpatrick}, {Light}, {Marsh}, {McCallon},
  {Schneider}, {Stiening}, {Sykes}, {Weinberg}, {Wheaton}, {Wheelock}, \&
  {Zacarias}}]{Cutri2003}
{Cutri}, R.~M., {Skrutskie}, M.~F., {van Dyk}, S., {et~al.} 2003, VizieR Online
  Data Catalog, II/246

\bibitem[{Faherty {et~al.}(2016)Faherty, Riedel, Cruz, Gagne, Filippazzo,
  Lambrides, Fica, Weinberger, Thorstensen, Tinney, Baldassare, Lemonier, \&
  Rice}]{Faherty2016}
Faherty, J.~K., Riedel, A.~R., Cruz, K.~L., {et~al.} 2016, The Astrophysical
  Journal Supplement Series, 225, 1

\bibitem[{Filippazzo {et~al.}(2015)Filippazzo, Rice, Faherty, Cruz, {Van
  Gordon}, \& Looper}]{Filippazzo2015}
Filippazzo, J.~C., Rice, E.~L., Faherty, J., {et~al.} 2015, The Astrophysical
  Journal, 810, 158

\bibitem[{Foreman-Mackey {et~al.}(2013)Foreman-Mackey, Hogg, Lang, \&
  Goodman}]{fm2013}
Foreman-Mackey, D., Hogg, D.~W., Lang, D., \& Goodman, J. 2013, Publications of
  the Astronomical Society of the Pacific, 125, 306

\bibitem[{{Freedman} {et~al.}(2014){Freedman}, {Lustig-Yaeger}, {Fortney},
  {Lupu}, {Marley}, \& {Lodders}}]{Freedman2014}
{Freedman}, R.~S., {Lustig-Yaeger}, J., {Fortney}, J.~J., {et~al.} 2014, \apjs,
  214, 25

\bibitem[{{Freedman} {et~al.}(2008){Freedman}, {Marley}, \&
  {Lodders}}]{Freedman2008}
{Freedman}, R.~S., {Marley}, M.~S., \& {Lodders}, K. 2008, \apjs, 174, 504

\bibitem[{{Gaarn} {et~al.}(2022){Gaarn}, {Burningham}, {Faherty}, Channon,
  Mark, {Gonzales}, {Calamari}, {Bardalez Gagliuffi}, {Lupu}, \&
  {Freedman}}]{Gaarn2022}
{Gaarn}, J., {Burningham}, B., {Faherty}, J.~K., {et~al.} 2022, \mnras

\bibitem[{Gagn{\'{e}} {et~al.}(2017)Gagn{\'{e}}, Faherty, Burgasser, Artigau,
  Bouchard, Albert, Lafreni{\`{e}}re, Doyon, \& Gagliuffi}]{Gagne2017}
Gagn{\'{e}}, J., Faherty, J.~K., Burgasser, A.~J., {et~al.} 2017, The
  Astrophysical Journal, 841, L1

\bibitem[{{Gaia Collaboration}(2018)}]{Gaia2018}
{Gaia Collaboration}. 2018, VizieR Online Data Catalog, I/345

\bibitem[{Gao {et~al.}(2020)Gao, Thorngren, Lee, Fortney, Morley, Wakeford,
  Powell, Stevenson, \& Zhang}]{Gao2020}
Gao, P., Thorngren, D.~P., Lee, G. K.~H., {et~al.} 2020, Nature Astronomy, 53,
  1689

\bibitem[{{Ge} {et~al.}(2019){Ge}, {Zhang}, {Fletcher}, {Orton}, {Sinclair},
  {Fernandes}, {Momary}, {Kasaba}, {Sato}, \& {Fujiyoshi}}]{Ge2019}
{Ge}, H., {Zhang}, X., {Fletcher}, L.~N., {et~al.} 2019, \aj, 157, 89

\bibitem[{{Gelino} \& {Marley}(2000)}]{Gelino2000}
{Gelino}, C., \& {Marley}, M. 2000, in Astronomical Society of the Pacific
  Conference Series, Vol. 212, From Giant Planets to Cool Stars, ed. C.~A.
  {Griffith} \& M.~S. {Marley}, 322

\bibitem[{{Gibbs} \& {Fitzgerald}(2022)}]{Gibbs2022}
{Gibbs}, A., \& {Fitzgerald}, M.~P. 2022, \aj, 164, 63

\bibitem[{{Gonzales} {et~al.}(2020){Gonzales}, {Burningham}, {Faherty},
  {Cleary}, {Visscher}, {Marley}, {Lupu}, \& {Freedman}}]{Gonzales2020}
{Gonzales}, E.~C., {Burningham}, B., {Faherty}, J.~K., {et~al.} 2020, \apj,
  905, 46

\bibitem[{{Gonzales} {et~al.}(2022){Gonzales}, {Burningham}, {Faherty},
  {Lewis}, {Visscher}, \& {Marley}}]{Gonzales2022}
---. 2022, \apj, 938, 56

\bibitem[{{Gonzales} {et~al.}(2021){Gonzales}, {Burningham}, {Faherty},
  {Visscher}, {Marley}, {Lupu}, {Freedman}, \& {Lewis}}]{Gonzales2021}
---. 2021, \apj, 923, 19

\bibitem[{Hallinan {et~al.}(2015)Hallinan, Littlefair, Cotter, Bourke, Harding,
  Pineda, Butler, Golden, Basri, Doyle, Kao, Berdyugina, Kuznetsov, Rupen, \&
  Antonova}]{Hallinan2015}
Hallinan, G., Littlefair, S.~P., Cotter, G., {et~al.} 2015, Nature, 523, 568

\bibitem[{{Hansen}(1971)}]{Hansen1971}
{Hansen}, J.~E. 1971, Journal of Atmospheric Sciences, 28, 1400

\bibitem[{{Helling} \& {Rimmer}(2019)}]{Helling2019}
{Helling}, C., \& {Rimmer}, P.~B. 2019, Philosophical Transactions of the Royal
  Society of London Series A, 377, 20180398

\bibitem[{{Helling} {et~al.}(2006){Helling}, {Thi}, {Woitke}, \&
  {Fridlund}}]{Helling2006}
{Helling}, C., {Thi}, W.~F., {Woitke}, P., \& {Fridlund}, M. 2006, \aap, 451,
  L9

\bibitem[{Helling {et~al.}(2008)Helling, Ackerman, Allard, Dehn, Hauschildt,
  Homeier, Lodders, Marley, Rietmeijer, Tsuji, \& Woitke}]{Helling2008}
Helling, C., Ackerman, A., Allard, F., {et~al.} 2008, Monthly Notices of the
  Royal Astronomical Society, 391, 1854

\bibitem[{{John}(1988)}]{John1988}
{John}, T.~L. 1988, \aap, 193, 189

\bibitem[{Kao {et~al.}(2018)Kao, Hallinan, Pineda, Stevenson, \&
  Burgasser}]{Kao2018}
Kao, M., Hallinan, G., Pineda, J.~S., Stevenson, D., \& Burgasser, A. 2018, The
  Astrophysical Journal Supplement Series, 237, 25

\bibitem[{Kao {et~al.}(2016)Kao, Hallinan, Pineda, Escala, Burgasser, Bourke,
  \& Stevenson}]{Kao2016}
Kao, M.~M., Hallinan, G., Pineda, J.~S., {et~al.} 2016, The Astrophysical
  Journal, 818, 24

\bibitem[{{Karalidi} {et~al.}(2021){Karalidi}, {Marley}, {Fortney}, {Morley},
  {Saumon}, {Lupu}, {Visscher}, \& {Freedman}}]{Karalidi2021}
{Karalidi}, T., {Marley}, M., {Fortney}, J.~J., {et~al.} 2021, \apj, 923, 269

\bibitem[{Kass \& Raftery(1995)}]{Kass1995}
Kass, R.~E., \& Raftery, A.~E. 1995, Journal of the American Statistical
  Association, 91, 477

\bibitem[{{Kitzmann} {et~al.}(2020){Kitzmann}, {Heng}, {Oreshenko}, {Grimm},
  {Apai}, {Bowler}, {Burgasser}, \& {Marley}}]{Kitzmann2020}
{Kitzmann}, D., {Heng}, K., {Oreshenko}, M., {et~al.} 2020, \apj, 890, 174

\bibitem[{{Lagrange} {et~al.}(2010){Lagrange}, {Bonnefoy}, {Chauvin}, {Apai},
  {Ehrenreich}, {Boccaletti}, {Gratadour}, {Rouan}, {Mouillet}, {Lacour}, \&
  {Kasper}}]{Lagrange2010}
{Lagrange}, A.~M., {Bonnefoy}, M., {Chauvin}, G., {et~al.} 2010, Science, 329,
  57

\bibitem[{{Lee} {et~al.}(2012){Lee}, {Fletcher}, \& {Irwin}}]{Lee2012}
{Lee}, J.~M., {Fletcher}, L.~N., \& {Irwin}, P.~G.~J. 2012, \mnras, 420, 170

\bibitem[{Lew {et~al.}(2016)Lew, Apai, Zhou, Schneider, Burgasser, Karalidi,
  Yang, Marley, Cowan, Bedin, Metchev, Radigan, \& Lowrance}]{Lew2016}
Lew, B. W.~P., Apai, D., Zhou, Y., {et~al.} 2016, The Astrophysical Journal,
  829, L32

\bibitem[{{Line} {et~al.}(2014){Line}, {Fortney}, {Marley}, \&
  {Sorahana}}]{Line2014}
{Line}, M.~R., {Fortney}, J.~J., {Marley}, M.~S., \& {Sorahana}, S. 2014, \apj,
  793, 33

\bibitem[{Line {et~al.}(2015)Line, Teske, Burningham, Fortney, \&
  Marley}]{Line2015}
Line, M.~R., Teske, J., Burningham, B., Fortney, J.~J., \& Marley, M.~S. 2015,
  Astrophysical Journal, 807, 183

\bibitem[{Line {et~al.}(2017)Line, Marley, Liu, Burningham, Morley, Hinkel,
  Teske, Fortney, Freedman, \& Lupu}]{Line2017}
Line, M.~R., Marley, M.~S., Liu, M.~C., {et~al.} 2017, The Astrophysical
  Journal, 848, 83

\bibitem[{Liu {et~al.}(2016)Liu, Dupuy, \& Allers}]{Liu2016}
Liu, M.~C., Dupuy, T.~J., \& Allers, K.~N. 2016, The Astrophysical Journal,
  833, 96

\bibitem[{{Lodders} \& {Fegley}(2002)}]{Lodders2002Icar}
{Lodders}, K., \& {Fegley}, B. 2002, \icarus, 155, 393

\bibitem[{{Lodders} \& {Fegley}(2006)}]{LoddersFegley2006}
{Lodders}, K., \& {Fegley}, B., J. 2006, in Astrophysics Update 2, ed. J.~W.
  {Mason}, 1

\bibitem[{{Lueber} {et~al.}(2022){Lueber}, {Kitzmann}, {Bowler}, {Burgasser},
  \& {Heng}}]{Lueber2022}
{Lueber}, A., {Kitzmann}, D., {Bowler}, B.~P., {Burgasser}, A.~J., \& {Heng},
  K. 2022, \apj, 930, 136

\bibitem[{{Luna} \& {Morley}(2021)}]{Luna2021}
{Luna}, J.~L., \& {Morley}, C.~V. 2021, \apj, 920, 146

\bibitem[{{MacDonald} \& {Lewis}(2022)}]{MacDonald2022}
{MacDonald}, R.~J., \& {Lewis}, N.~K. 2022, \apj, 929, 20

\bibitem[{{MacDonald} \& {Madhusudhan}(2017)}]{MacDonald2017}
{MacDonald}, R.~J., \& {Madhusudhan}, N. 2017, \mnras, 469, 1979

\bibitem[{Macintosh {et~al.}(2015)Macintosh, Graham, Barman, {De Rosa},
  Konopacky, Marley, Marois, Nielsen, Pueyo, Rajan, Rameau, Saumon, Wang,
  Patience, Ammons, Arriaga, Artigau, Beckwith, Brewster, Bruzzone, Bulger,
  Burningham, Burrows, Chen, Chiang, Chilcote, Dawson, Dong, Doyon, Draper,
  Duchene, Esposito, Fabrycky, Fitzgerald, Follette, Fortney, Gerard, Goodsell,
  Greenbaum, Hibon, Hinkley, Cotten, Hung, Ingraham, Johnson-Groh, Kalas,
  Lafreniere, Larkin, Lee, Line, Long, Maire, Marchis, Matthews, Max, Metchev,
  Millar-Blanchaer, Mittal, Morley, Morzinski, Murray-Clay, Oppenheimer,
  Palmer, Patel, Perrin, Poyneer, Rafikov, Rantakyro, Rice, Rojo, Rudy, Ruffio,
  Ruiz, Sadakuni, Saddlemyer, Salama, Savransky, Schneider, Sivaramakrishnan,
  Song, Soummer, Thomas, Vasisht, Wallace, Ward-Duong, Wiktorowicz, Wolff, \&
  Zuckerman}]{Macintosh2015}
Macintosh, B., Graham, J.~R., Barman, T., {et~al.} 2015, Science, 350, 64

\bibitem[{{Madhusudhan}(2012)}]{Madhusudhan2012}
{Madhusudhan}, N. 2012, \apj, 758, 36

\bibitem[{Madhusudhan \& Seager(2009)}]{Madhusudhan2009}
Madhusudhan, N., \& Seager, S. 2009, Astrophysical Journal, 707, 24

\bibitem[{Marley {et~al.}(1996)Marley, Saumon, Guillot, Freedman, Hubbard,
  Burrows, \& Lunine}]{Marley1996}
Marley, M.~S., Saumon, D., Guillot, T., {et~al.} 1996, Science, 272, 1919

\bibitem[{{Marley} {et~al.}(2021){Marley}, {Saumon}, {Visscher}, {Lupu},
  {Freedman}, {Morley}, {Fortney}, {Seay}, {Smith}, {Teal}, \&
  {Wang}}]{Marley2021}
{Marley}, M.~S., {Saumon}, D., {Visscher}, C., {et~al.} 2021, \apj, 920, 85

\bibitem[{Marois {et~al.}(2008)Marois, Macintosh, Barman, Zuckerman, Song,
  Patience, Lafreniere, \& Doyon}]{Marois2008}
Marois, C., Macintosh, B., Barman, T.~S., {et~al.} 2008, Science, 322, 1348

\bibitem[{{McKay} {et~al.}(1989){McKay}, {Pollack}, \& {Courtin}}]{McKay1989}
{McKay}, C.~P., {Pollack}, J.~B., \& {Courtin}, R. 1989, \icarus, 80, 23

\bibitem[{Metchev {et~al.}(2015)Metchev, Heinze, Apai, Flateau, Radigan,
  Burgasser, Marley, Artigau, Plavchan, \& Goldman}]{Metchev2015}
Metchev, S.~A., Heinze, A., Apai, D., {et~al.} 2015, The Astrophysical Journal,
  799, 154

\bibitem[{{Miles} {et~al.}(2022){Miles}, {Biller}, {Patapis}, {Worthen},
  {Rickman}, {Hoch}, {Skemer}, {Perrin}, {Chen}, {Mukherjee}, {Morley},
  {Moran}, {Bonnefoy}, {Petrus}, {Carter}, {Choquet}, {Hinkley}, {Ward-Duong},
  {Leisenring}, {Millar-Blanchaer}, {Pueyo}, {Ray}, {Stapelfeldt}, {Stone},
  {Wang}, {Absil}, {Balmer}, {Boccaletti}, {Bonavita}, {Booth}, {Bowler},
  {Chauvin}, {Christiaens}, {Currie}, {Danielski}, {Fortney}, {Girard},
  {Greenbaum}, {Henning}, {Hines}, {Janson}, {Kalas}, {Kammerer}, {Kenworthy},
  {Kervella}, {Lagage}, {Lew}, {Liu}, {Macintosh}, {Marino}, {Marley},
  {Marois}, {Matthews}, {Matthews}, {Mawet}, {McElwain}, {Metchev}, {Meyer},
  {Molliere}, {Pantin}, {Rebollido}, {Ren}, {Vasist}, {Wyatt}, {Zhou},
  {Briesemeister}, {Bryan}, {Calissendorff}, {Catalloube}, {Cugno}, {De Furio},
  {Dupuy}, {Factor}, {Faherty}, {Fitzgerald}, {Franson}, {Gonzales}, {Hood},
  {Howe}, {Kraus}, {Kuzuhara}, {Lawson}, {Lazzoni}, {Liu}, {Llop-Sayson},
  {Lloyd}, {Martinez}, {Mazoyer}, {Quanz}, {Adams Redai}, {Samland},
  {Schlieder}, {Tamura}, {Tan}, {Uyama}, {Vigan}, {Vos}, {Wagner}, {Wolff},
  {Ygouf}, {Zhang}, \& {Zhang}}]{Miles2022}
{Miles}, B.~E., {Biller}, B.~A., {Patapis}, P., {et~al.} 2022, arXiv e-prints,
  arXiv:2209.00620

\bibitem[{{Molli{\`e}re} {et~al.}(2019){Molli{\`e}re}, {Wardenier}, {van
  Boekel}, {Henning}, {Molaverdikhani}, \& {Snellen}}]{Molliere2019}
{Molli{\`e}re}, P., {Wardenier}, J.~P., {van Boekel}, R., {et~al.} 2019, \aap,
  627, A67

\bibitem[{{Molli{\`e}re} {et~al.}(2020){Molli{\`e}re}, {Stolker}, {Lacour},
  {Otten}, {Shangguan}, {Charnay}, {Molyarova}, {Nowak}, {Henning}, {Marleau},
  {Semenov}, {van Dishoeck}, {Eisenhauer}, {Garcia}, {Garcia Lopez}, {Girard},
  {Greenbaum}, {Hinkley}, {Kervella}, {Kreidberg}, {Maire}, {Nasedkin},
  {Pueyo}, {Snellen}, {Vigan}, {Wang}, {de Zeeuw}, \& {Zurlo}}]{Molliere2020}
{Molli{\`e}re}, P., {Stolker}, T., {Lacour}, S., {et~al.} 2020, \aap, 640, A131

\bibitem[{{Molli{\`e}re} {et~al.}(2022){Molli{\`e}re}, {Molyarova}, {Bitsch},
  {Henning}, {Schneider}, {Kreidberg}, {Eistrup}, {Burn}, {Nasedkin},
  {Semenov}, {Mordasini}, {Schlecker}, {Schwarz}, {Lacour}, {Nowak}, \&
  {Schulik}}]{Molliere2022}
{Molli{\`e}re}, P., {Molyarova}, T., {Bitsch}, B., {et~al.} 2022, \apj, 934, 74

\bibitem[{Morley {et~al.}(2012)Morley, Fortney, Marley, Visscher, Saumon, \&
  Leggett}]{Morley2012}
Morley, C.~V., Fortney, J.~J., Marley, M.~S., {et~al.} 2012, The Astrophysical
  Journal, 756, 172

\bibitem[{{Nissen}(2013)}]{Nissen2013}
{Nissen}, P.~E. 2013, \aap, 552, A73

\bibitem[{{Nissen}(2015)}]{Nissen2015}
---. 2015, \aap, 579, A52

\bibitem[{{Nixon} \& {Madhusudhan}(2022)}]{Nixon2022}
{Nixon}, M.~C., \& {Madhusudhan}, N. 2022, \apj, 935, 73

\bibitem[{{{\"O}berg} {et~al.}(2011){{\"O}berg}, {Murray-Clay}, \&
  {Bergin}}]{Oberg2011}
{{\"O}berg}, K.~I., {Murray-Clay}, R., \& {Bergin}, E.~A. 2011, \apjl, 743, L16

\bibitem[{{Perry} {et~al.}(1999){Perry}, {Kim}, {Fox}, \& {Porter}}]{Perry1999}
{Perry}, J.~J., {Kim}, Y.~H., {Fox}, J.~L., \& {Porter}, H.~S. 1999, \jgr, 104,
  16541

\bibitem[{Pineda {et~al.}(2017)Pineda, Hallinan, \& Kao}]{Pineda2017}
Pineda, J.~S., Hallinan, G., \& Kao, M.~M. 2017, The Astrophysical Journal,
  846, 75

\bibitem[{{Pineda} {et~al.}(2016){Pineda}, {Hallinan}, {Kirkpatrick}, {Cotter},
  {Kao}, \& {Mooley}}]{Pineda2016}
{Pineda}, J.~S., {Hallinan}, G., {Kirkpatrick}, J.~D., {et~al.} 2016, \apj,
  826, 73

\bibitem[{Radigan {et~al.}(2012)Radigan, Jayawardhana, Lafreni{\`{e}}re,
  Artigau, Marley, \& Saumon}]{Radigan2012}
Radigan, J., Jayawardhana, R., Lafreni{\`{e}}re, D., {et~al.} 2012, The
  Astrophysical Journal, 750, 105

\bibitem[{Radigan {et~al.}(2014)Radigan, Lafreni{\`{e}}re, Jayawardhana, \&
  Artigau}]{Radigan2014}
Radigan, J., Lafreni{\`{e}}re, D., Jayawardhana, R., \& Artigau, E. 2014, The
  Astrophysical Journal, 793, 75

\bibitem[{{Richard} {et~al.}(2012){Richard}, {Gordon}, {Rothman}, {Abel},
  {Frommhold}, {Gustafsson}, {Hartmann}, {Hermans}, {Lafferty}, {Orton},
  {Smith}, \& {Tran}}]{Richard2012}
{Richard}, C., {Gordon}, I.~E., {Rothman}, L.~S., {et~al.} 2012, \jqsrt, 113,
  1276

\bibitem[{{Richey-Yowell} {et~al.}(2020){Richey-Yowell}, {Kao}, {Pineda},
  {Shkolnik}, \& {Hallinan}}]{Richey-Yowell2020}
{Richey-Yowell}, T., {Kao}, M.~M., {Pineda}, J.~S., {Shkolnik}, E.~L., \&
  {Hallinan}, G. 2020, \apj, 903, 74

\bibitem[{Saumon \& Marley(2008)}]{Saumon2008}
Saumon, D., \& Marley, M. 2008, The Astrophysical Journal, 689, 1327

\bibitem[{{Saumon} {et~al.}(2012){Saumon}, {Marley}, {Abel}, {Frommhold}, \&
  {Freedman}}]{Saumon2012}
{Saumon}, D., {Marley}, M.~S., {Abel}, M., {Frommhold}, L., \& {Freedman},
  R.~S. 2012, \apj, 750, 74

\bibitem[{Schwarz(1978)}]{Schwarz1978}
Schwarz, G. 1978, The Annals of Statistics, 6, 461

\bibitem[{{Smart} {et~al.}(2013){Smart}, {Tinney}, {Bucciarelli}, {Marocco},
  {Abbas}, {Andrei}, {Bernardi}, {Burningham}, {Cardoso}, {Costa}, {Crosta},
  {Dapr{\'a}}, {Day-Jones}, {Goldman}, {Jones}, {Lattanzi}, {Leggett}, {Lucas},
  {Mendez}, {Penna}, {Pinfield}, {Smith}, {Sozzetti}, \&
  {Vecchiato}}]{Smart2013}
{Smart}, R.~L., {Tinney}, C.~G., {Bucciarelli}, B., {et~al.} 2013, \mnras, 433,
  2054

\bibitem[{{Sorahana} {et~al.}(2013){Sorahana}, {Yamamura}, \&
  {Murakami}}]{Sorahana2013}
{Sorahana}, S., {Yamamura}, I., \& {Murakami}, H. 2013, \apj, 767, 77

\bibitem[{{Su{\'a}rez} \& {Metchev}(2022)}]{Suarez2022}
{Su{\'a}rez}, G., \& {Metchev}, S. 2022, \mnras, 513, 5701

\bibitem[{{Tan} \& {Showman}(2021)}]{Tan2021}
{Tan}, X., \& {Showman}, A.~P. 2021, \mnras, 502, 678

\bibitem[{{Toon} {et~al.}(1989){Toon}, {McKay}, {Ackerman}, \&
  {Santhanam}}]{Toon1989}
{Toon}, O.~B., {McKay}, C.~P., {Ackerman}, T.~P., \& {Santhanam}, K. 1989,
  \jgr, 94, 16287

\bibitem[{{Tsuji}(2002)}]{Tsuji2002}
{Tsuji}, T. 2002, \apj, 575, 264

\bibitem[{{Vasavada} {et~al.}(1999){Vasavada}, {Bouchez}, {Ingersoll},
  {Little}, {Anger}, \& {Galileo SSI Team}}]{Vasavada1999}
{Vasavada}, A.~R., {Bouchez}, A.~H., {Ingersoll}, A.~P., {et~al.} 1999, \jgr,
  104, 27133

\bibitem[{Visscher {et~al.}(2010)Visscher, Lodders, \& Fegley}]{Visscher2010}
Visscher, C., Lodders, K., \& Fegley, B. 2010, The Astrophysical Journal, 716,
  1060

\bibitem[{Vos {et~al.}(2017)Vos, Allers, \& Biller}]{Vos2017}
Vos, J.~M., Allers, K.~N., \& Biller, B.~A. 2017, The Astrophysical Journal,
  842, 78

\bibitem[{{Vos} {et~al.}(2022){Vos}, {Faherty}, {Gagn{\'e}}, {Marley},
  {Metchev}, {Gizis}, {Rice}, \& {Cruz}}]{Vos2022}
{Vos}, J.~M., {Faherty}, J.~K., {Gagn{\'e}}, J., {et~al.} 2022, \apj, 924, 68

\bibitem[{Vos {et~al.}(2019)Vos, Biller, Bonavita, Eriksson, Liu, Best,
  Metchev, Radigan, Allers, Janson, Buenzli, Dupuy, Bonnefoy, Manjavacas,
  Brandner, Crossfield, Deacon, Henning, Homeier, Kopytova, \&
  Schlieder}]{Vos2019}
Vos, J.~M., Biller, B.~A., Bonavita, M., {et~al.} 2019, Monthly Notices of the
  Royal Astronomical Society, 483, 480

\bibitem[{Yang {et~al.}(2016)Yang, Apai, Marley, Karalidi, Flateau, Showman,
  Metchev, Buenzli, Radigan, Artigau, Lowrance, \& Burgasser}]{Yang2016}
Yang, H., Apai, D., Marley, M.~S., {et~al.} 2016, The Astrophysical Journal,
  826, 8

\bibitem[{Zalesky {et~al.}(2019)Zalesky, Line, Schneider, \&
  Patience}]{Zalesky2019}
Zalesky, J.~A., Line, M.~R., Schneider, A.~C., \& Patience, J. 2019, The
  Astrophysical Journal, 877, 24

\bibitem[{{Zhang} {et~al.}(2021{\natexlab{a}}){Zhang}, {Snellen}, \&
  {Molli{\`e}re}}]{ZhangSnellen2021}
{Zhang}, Y., {Snellen}, I. A.~G., \& {Molli{\`e}re}, P. 2021{\natexlab{a}},
  \aap, 656, A76

\bibitem[{{Zhang} {et~al.}(2021{\natexlab{b}}){Zhang}, {Liu}, {Best}, {Dupuy},
  \& {Siverd}}]{Zhang2021}
{Zhang}, Z., {Liu}, M.~C., {Best}, W. M.~J., {Dupuy}, T.~J., \& {Siverd}, R.~J.
  2021{\natexlab{b}}, \apj, 911, 7

\bibitem[{Zhou {et~al.}(2016)Zhou, Apai, Schneider, Marley, \&
  Showman}]{Zhou2016}
Zhou, Y., Apai, D., Schneider, G.~H., Marley, M.~S., \& Showman, A.~P. 2016,
  The Astrophysical Journal, 818, 176

\bibitem[{{Zhou} {et~al.}(2020){Zhou}, {Bowler}, {Morley}, {Apai}, {Kataria},
  {Bryan}, \& {Benneke}}]{Zhou2020}
{Zhou}, Y., {Bowler}, B.~P., {Morley}, C.~V., {et~al.} 2020, \aj, 160, 77

\bibitem[{{Zhou} {et~al.}(2019){Zhou}, {Apai}, {Lew}, {Schneider},
  {Manjavacas}, {Bedin}, {Cowan}, {Marley}, {Radigan}, {Karalidi}, {Lowrance},
  {Miles-P{\'a}ez}, {Metchev}, \& {Burgasser}}]{Zhou2019}
{Zhou}, Y., {Apai}, D., {Lew}, B. W.~P., {et~al.} 2019, \aj, 157, 128

\end{thebibliography}
\bibliographystyle{aasjournal}

\restartappendixnumbering
\appendix
\section{Corner plots}

\begin{figure*}
   \centering
   \includegraphics[scale=.22]{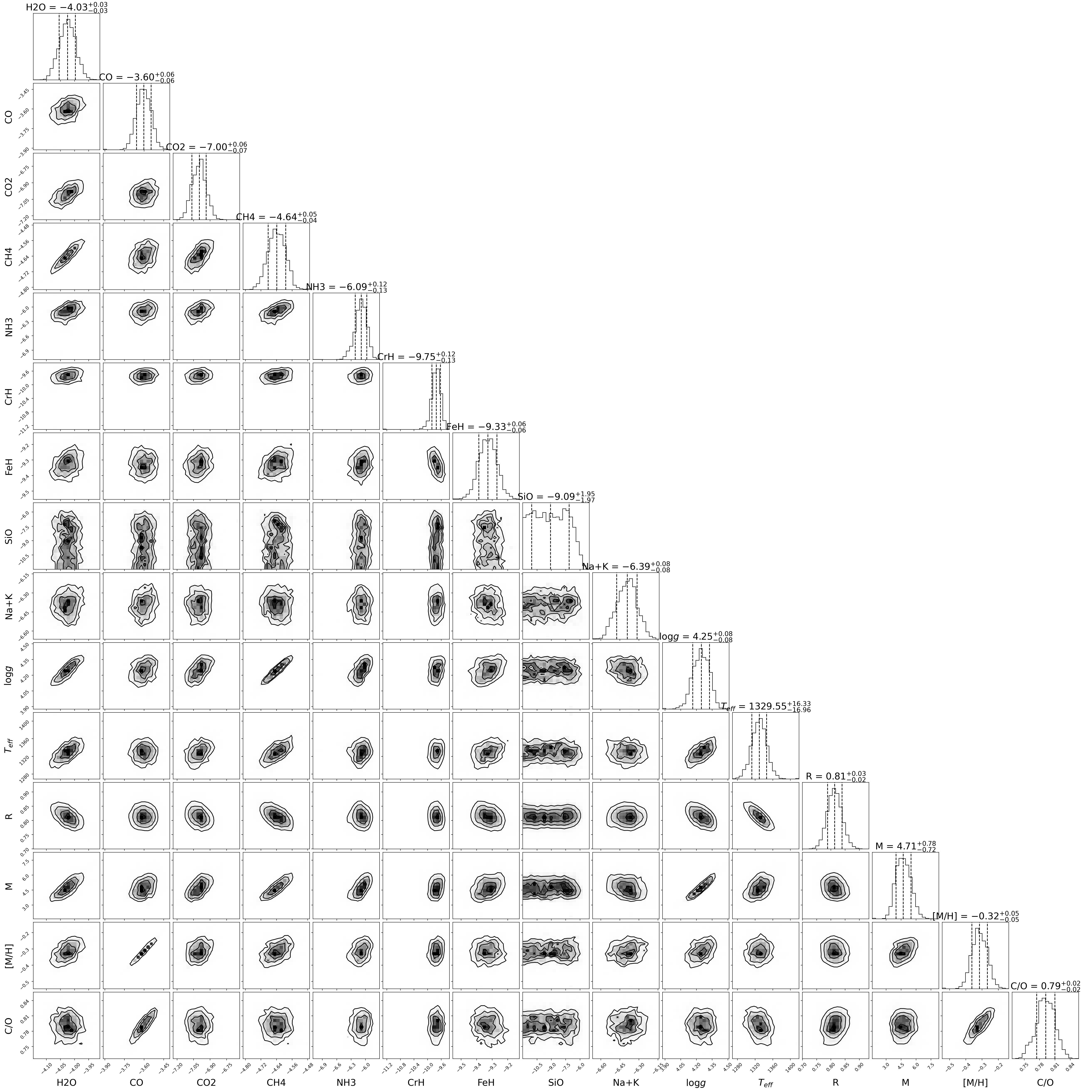}
   \caption{Corner plot showing the gas opacities, $\log g$ and derived $T_{\mathrm{eff}}$, $R$, $M$, [M/H] and C/O ratio for the top-ranked model for \obj{s0136}.}
   \label{fig:0136_gascorner}
\end{figure*}

\begin{figure*}
   \centering
   \includegraphics[scale=.22]{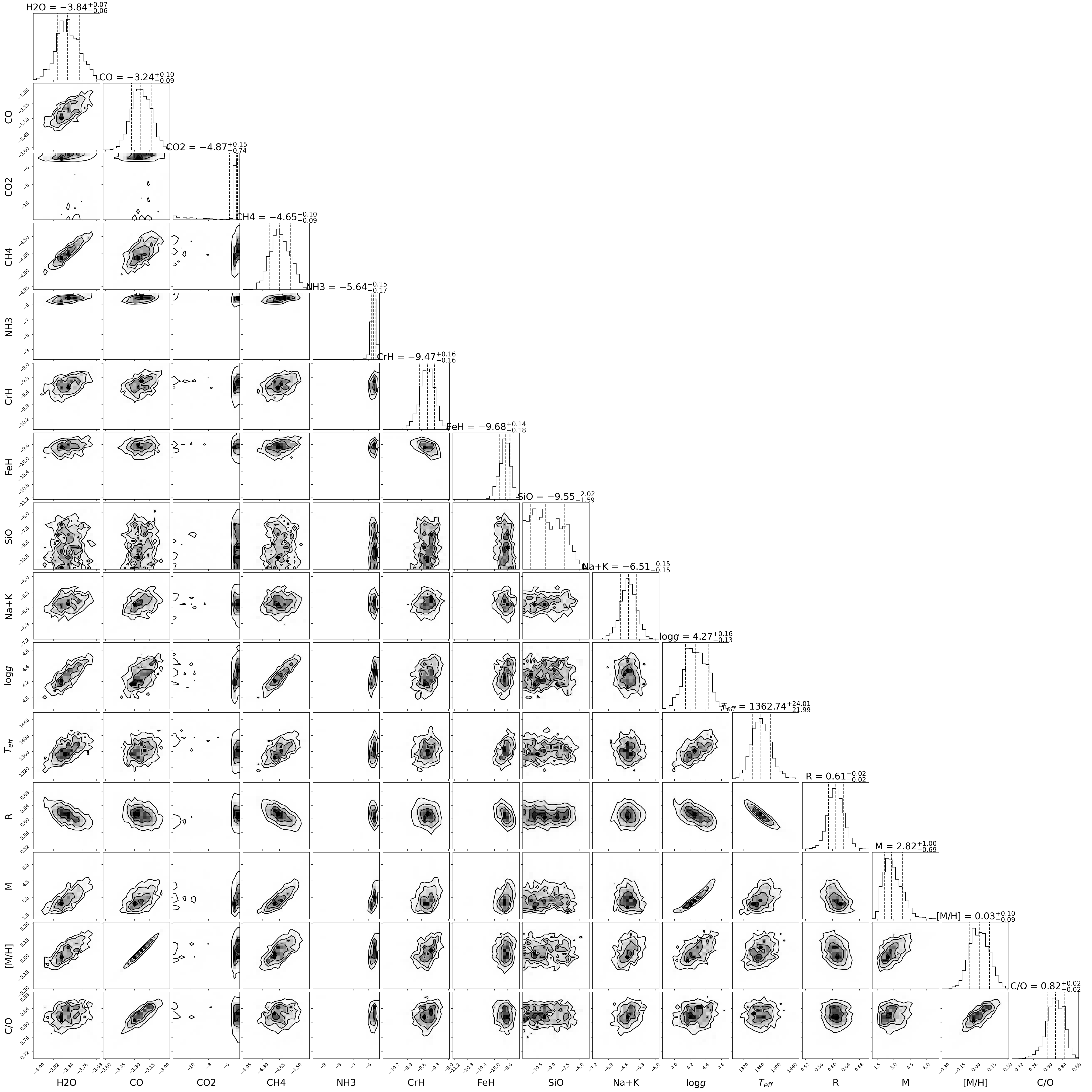}
   \caption{Corner plot showing the gas opacities, $\log g$ and derived $T_{\mathrm{eff}}$, $R$, $M$, [M/H] and C/O ratio for the top-ranked model for \obj{2m2139}. }
   \label{fig:2139_gascorner}
\end{figure*}

\end{document}